\documentclass[numberedappendix]{emulateapj}
\usepackage{graphicx, hyphenat, amsmath, color}
\graphicspath{ {/Users/salberts/xxx/} }

\providecommand{\e}[1]{\ensuremath{\times 10^{#1}}}
\newcommand{\Msun}{\ensuremath{\mathrm{M_{\odot}}}}
\newcommand{\Lsun}{\ensuremath{\mathrm{L_{\odot}}}}

\def\lesssim{\mathrel{\hbox{\rlap{\hbox{\lower3pt\hbox{$\sim$}}}\hbox{\raise2pt\hbox{$<$}}}}}
\def\gtrsim{\mathrel{\hbox{\rlap{\hbox{\lower3pt\hbox{$\sim$}}}\hbox{\raise2pt\hbox{$>$}}}}}

\shorttitle{Star Formation and AGN in $z=1-2$ Galaxy Clusters}
\shortauthors{Alberts et al.}
\slugcomment{{\sc Accepted to ApJ:} April 10, 2016} 

\begin{document}

\title{Star Formation and AGN Activity in Galaxy Clusters from z=1-2: a Multi-wavelength Analysis Featuring Herschel/PACS}
\author{Stacey Alberts\altaffilmark{1,2}, Alexandra Pope\altaffilmark{2}, Mark Brodwin\altaffilmark{3}, Sun Mi Chung\altaffilmark{4}, Ryan Cybulski\altaffilmark{2}, Arjun Dey\altaffilmark{5}, Peter R. M. Eisenhardt\altaffilmark{6}, Audrey Galametz\altaffilmark{7}, Anthony H. Gonzalez\altaffilmark{8}, Buell T. Jannuzi\altaffilmark{1}, S. Adam Stanford\altaffilmark{9}, Gregory F. Snyder\altaffilmark{10}, Daniel Stern\altaffilmark{6}, Gregory R. Zeimann\altaffilmark{11}}
\altaffiltext{1}{Steward Observatory, University of Arizona, 933 North Cherry Avenue, Tucson, AZ 85721, USA}
\altaffiltext{2}{Department of Astronomy, University of Massachusetts, LGRT-B 619E, Amherst, MA 01003, USA}
\altaffiltext{3}{Department of Physics and Astronomy, University of Missouri, 5110 Rockhill Road, Kansas City, MO 64110}
\altaffiltext{4}{Department of Astronomy, The Ohio State University, 140 W 18th Avenue, Columbus, OH 43210, USA}
\altaffiltext{5}{National Optical Astronomy Observatory, 950 N. Cherry Ave., Tucson, AZ 85719}
\altaffiltext{6}{Jet Propulsion Laboratory, California Institute of Technology, Pasadena, CA 91109}
\altaffiltext{7}{Max-Planck-Institut fuer Extraterrestrische Physik, Giessenbachstrasse, D-85748 Garching, Germany}
\altaffiltext{8}{Department of Astronomy, University of Florida, Gainesville, FL 32611-2055}
\altaffiltext{9}{Physics Department, One Shields Avenue, University of California, Davis, CA 95616, USA}
\altaffiltext{10}{Space Telescope Science Institute, 3700 San Martin Dr, Baltimore, MD 21218, USA}
\altaffiltext{11}{Department of Astronomy and Astrophysics, Pennsylvania State University, 525 Davey Laboratory, University Park, Pennsylvania 16802}

% Abstract should be < 250 words for ApJ

\begin{abstract}
We present a detailed, multi-wavelength study of star formation (SF) and AGN activity  in 11 near-infrared (IR) selected, spectroscopically confirmed, massive ($\gtrsim10^{14}\,\Msun$) galaxy clusters at $1<z<1.75$.  Using new, deep {\it Herschel}/PACS imaging, we characterize the optical to far-IR spectral energy distributions (SEDs) for IR-luminous cluster galaxies, finding that they can, on average, be well described by field galaxy templates.  Identification and decomposition of AGN through SED fittings allows us to include the contribution to cluster SF from AGN host galaxies.  We quantify the star-forming fraction, dust-obscured SF rates (SFRs), and specific-SFRs for cluster galaxies as a function of cluster-centric radius and redshift.    In good agreement with previous studies, we find that SF in cluster galaxies at $z\gtrsim1.4$ is largely consistent with field galaxies at similar epochs, indicating an era before significant quenching in the cluster cores ($r<0.5\,$Mpc).  This is followed by a transition to lower SF activity as environmental quenching dominates by $z\sim1$.  Enhanced SFRs are found in lower mass ($10.1< \log \rm{M_{\star}}/\Msun<10.8$) cluster galaxies. We find significant variation in SF from cluster-to-cluster within our uniformly selected sample, indicating that caution should be taken when evaluating individual clusters.  We examine AGN in clusters from $z=0.5-2$, finding an excess AGN fraction at $z\gtrsim1$, suggesting environmental triggering of AGN during this epoch.  We argue that our results  $-$ a transition from field-like to quenched SF, enhanced SF in lower mass galaxies in the cluster cores, and excess AGN $-$ are consistent with a co-evolution between SF and AGN in clusters and an increased merger rate in massive haloes at high redshift.

\end{abstract}


\section{Introduction}
\label{sec:intro}

A complete understanding of galaxy evolution requires describing galaxy properties in relation to their environment over cosmic time.  Locally, galaxy populations show a strong anti-correlation between star formation rate (SFR) and surrounding galaxy density, with the cores of massive galaxy clusters inhabited by passively evolving early-type galaxies (ETGs).  Star forming galaxies (SFGs), on the other hand, reside preferentially in regions of lower galaxy density, such as groups or the field \citep[e.g.][]{dre80}.  Environmental quenching in the local Universe has been found to be highly efficient, observed as far from cluster centers as three times the virial radius \citep{chu11}, with infalling galaxies and groups experiencing pre-processing as they fall into massive haloes \citep[e.g.][]{bai09, chu10, cyb14}.

Beyond the local Universe, studies of galaxy clusters as a function of redshift are providing key insights into the environment's role in transforming SFGs into today's passive cluster populations. Direct observations of star formation (SF) in clusters up to $z\sim1$ have found a rapid evolution in cluster galaxy properties, with the fractions of Luminous Infrared Galaxies (LIRGs; $10^{11}\,\Lsun<{\rm L_{\rm IR}}<10^{12}\,\Lsun$) and Ultra-Luminous Infrared Galaxies (ULIRGs; L$_{\rm IR}>10^{12}\,\Lsun$) steadily increasing in cluster environments during this epoch \citep{coi05, gea06, mar07, muz08, koy08, hai09, fin10, smi10, chu11,web13}.  Measurements of the integrated SFR per unit halo mass in clusters have found that this quantity is evolving as fast or faster than in the field, $(1+z)^{5-7}$ up to $z\sim1$  \citep[e.g.][]{sai08, bai09, web13, hai13, pop14}.  Despite this evolution in star forming populations, however, dense cluster cores are still characterized by significant quenching up to $z\sim1$ \citep[e.g.][]{pat09,fin10,vul10,muz12}, consistent with the local SFR-density relation.  Optical and near-infrared (NIR) analyses of the colors and luminosity functions of cluster galaxies at $z<1$ favor cluster formation models with high formation redshifts ($z\gtrsim2-3$), in which clusters form in a burst of intense star formation activity and then largely passively evolve to $z\sim0$ \citep[e.g.][]{sta98, bla06, eis08, mei09}.  

It is only recently that studies of clusters at $z=1-2$ have begun to paint a different picture. Evidence for a departure from passive evolution models in massive haloes at $z>1$ was presented in \citet{man10}, which demonstrated that the 3.6 and 4.5$\,\mu$m luminosity functions of cluster galaxies from the IRAC Shallow Cluster Survey \citep[ISCS;][]{eis08} indicate significant rapid mass assembly at $z\gtrsim1.3$, much lower than the expected formation redshift. Studies of star formation in individual clusters at $z>1$ have found multiple examples of SFG populations in cluster cores, suggesting that the local SFR-density relation no longer holds at this epoch \citep[e.g.][]{tra10, hil10, hay11, fas11, tad12, zei13, bay13, fas14, san14, mei15, san15, ma15}.  It should be noted, however, that examples of seemingly evolved systems, with the local relation still in place, have also been found at high redshift \citep[$z\gtrsim1.5$;][]{koy14,new14} and that strong variations in galaxy populations can exist from cluster-to-cluster even at the same epoch \citep[e.g.][]{gea06, bro13}, making it a challenge to place individual clusters in the broader context.  This challenge is further compounded by different cluster selection techniques and different methods of measuring cluster galaxy properties, pointing to the need for studies of uniformly selected, statistical cluster samples. 

Using 16 spectroscopically confirmed clusters at $z>1$ from the ISCS, \citet{bro13} identified a transition epoch from an era of unquenched star formation in the cores of clusters at $z\sim1.4$ to significant quenching by $z\sim1$.  A {\it Herschel}/SPIRE stacking analysis of stellar mass-limited cluster galaxy samples for $\sim300$ ISCS clusters from $0.5<z<1.5$ demonstrated that cluster populations at $z\gtrsim1.4$ have, on average, SFRs and specific-SFRs (SSFRs) consistent with the field, followed by rapid evolution toward quenched SF at lower redshifts \citep{alb14}.    This decline in SF in clusters over cosmic time is paralleled by a decrease in black hole activity, with the fraction of active galactic nuclei (AGN) falling by two orders of magnitude in clusters from $z\sim1.5$ to $z\sim0$ \citep{gal09,mar09, mar13}.   These recent results indicate that the period from $z=1-2$ is pivotal for the mass assembly of cluster galaxies and our understanding of SF activity and galaxy evolution as a function of environment.  This epoch roughly coincides with the peak in the global SFR density of the Universe \citep[$z\sim1-3$;][]{mur11a, mag13} as well as the peak in the black hole growth in galaxies \citep[e.g.][]{sil08}.  During this time, the majority of the light from SF is enshrouded by dust and re-emitted in the infrared \citep[e.g.][]{mur11a}.

In this study, we present new, deep {\it Herschel}/PACS imaging of 11 massive ($\ga10^{14}\,\Msun$) galaxy clusters at $z=1-1.75$ from the ISCS and IRAC Distant Cluster Survey \citep[IDCS;][]{sta12}.  Combining this new infrared data with previously existing observations, we quantify the average optical-to-FIR SEDs of cluster galaxies at $z>1$. Probing near the peak of the dust emission of the spectral energy distribution (SED; 80-36$\,\mu$m over $z=1-1.75$) allows us to quantify robust dust-obscured SFRs in cluster galaxies selected using spectroscopic and photometric redshifts.  We further identify AGN emission in cluster galaxies through SED fitting \citep[e.g.][]{assef10, chu14} using extensive multi-wavelength photometry in the X-ray to the mid-infrared (MIR).  This information is used to account for the contribution to the cluster SF from AGN host galaxies in our 11 clusters with PACS imaging as well as examine the AGN fraction as a function of environment in $\sim250$ clusters from the ISCS/IDCS over the redshift range $z=0.5-2$.

In Section 2, we provide details of our cluster sample, observations, spectroscopic and photometric redshifts, and describe our procedure for identifying cluster members.  In Section 3, we present the SF properties of PACS-selected cluster members, including optical-to-FIR SEDs of cluster SFGs and AGN hosts.  Based on the cluster galaxy SEDs, we select appropriate templates and calculate robust SFRs and SSFRs for the IR-luminous galaxy (i.e. galaxies bright in the far-IR) population of clusters as a function of cluster-centric radius and redshift.  This analysis is then expanded to stellar mass-limited cluster galaxy samples using stacking.  The variation in total SF from cluster-to-cluster and the integrated SFR per unit halo mass as a function of redshift are presented.   Finally, we utilize clusters from the ISCS/IDCS over the redshift range $0.5<z<2$ to trace the evolution of the AGN fraction in galaxies as a function of environment and redshift.  Section 4 presents our discussion and Section 5 our summary and conclusions.  Throughout this work,  we adopt concordance cosmology: ($\Omega_{\Lambda}, \Omega_{M}, h$)=(0.7, 0.3, 0.7).  A \citet{kro01} IMF  is assumed unless otherwise stated.

\section{Data}
\label{sec:data}

\subsection{IRAC Shallow and Distant Cluster Surveys}
\label{sec:survey}

The ISCS consists of over 300 galaxy cluster candidates over the redshift range $0.1<z<2$.  The cluster candidates are identified based on infrared-selected galaxy catalogs as 3-D overdensities in (RA, Dec, photometric redshift) space using a wavelet detection algorithm and photometric redshifts  \citep{els06, bro06, eis08, bro13} derived from deep optical $B_wRI$ imaging from the NOAO Deep Wide-Field Survey \citep[NDWFS;][]{jan99} and {\it Spitzer}/IRAC imaging from the IRAC Shallow Survey \citep[ISS;][]{eis04}. The ISCS spans 8.5 square degrees in the Bo\"{o}tes field and includes $>100$ cluster candidates at $z>1$, over 20 of which have been spectroscopically confirmed \citep{sta05,bro06,bro11,bro13,els06,eis08,zei13}.  A follow-up survey, the IDCS, was conducted using deeper IRAC data from the {\it Spitzer} Deep, Wide-Field Survey \citep[SDWFS;][]{ash09}.  Two $z\sim1.8$ IDCS clusters have been spectroscopically confirmed to date \citep{sta12, zei12}.

%{\bf Given the expected cluster mass function and the depth of the ISS (8.8$\,\mu$Jy, 5$\sigma$ at 4.5$\,\mu$m), the ISCS cluster sample is essentially mass selected, with a typical halo mass of $\sim\,10^{14}\,\Msun$.  
Targeted follow-up of high significance, spectroscopically confirmed ISCS clusters at $z>1$ have found halo masses in the range M$_{200}$ = $(1-5)\times10^{14}\,\Msun$ using X-ray observations \citep{bro11,bro15} and weak lensing\citep{jee11}.  These direct measurements are consistent with statistical analyses of the full ISCS sample, which find mean halo masses of M$_{200}\sim(5-8)\times10^{13}\,\Msun$ using clustering \citep{bro07} and halo mass ranking simulations \citep{lin13}, with no significant redshift evolution in the median halo mass \citep{alb14}.  Given these typical halo masses, the ISCS clusters have a characteristic virial radius of  $\sim1\,$Mpc at $z>0.5$, which we will adopt for $r_{200}$ throughout this study. Though this work will primarily focus on ISCS clusters, we additionally include in our study one cluster at $z=1.75$ from the IDCS, which has a halo mass of M$_{200} \approx 4\times10^{14}\,\Msun$ from X-ray and Sunyaev-Zel'dovich effect measurements \citep{sta12, bro12, bro15}.

In this work, we concentrate our analysis on 11 spectroscopically confirmed clusters from the ISCS/IDCS that we observed with {\it Herschel}/PACS.  These clusters, which span the redshift range $1<z<1.75$, are listed in Table~\ref{tbl:clusters}, which includes available halo mass measurements and additional references.  In Section~\ref{sec:agn}, we utilize $\sim250$ ISCS clusters at $z>0.5$ plus three IDCS clusters at $z\sim1.8$ for an analysis of the AGN fraction in clusters. 

\begin{deluxetable*}{cccccccc}
\tabletypesize{\footnotesize}
\tablecolumns{8}
\tablewidth{0pt}
\tablecaption{Cluster Sample with Deep {\it Herschel}/PACS Imaging \label{tbl:clusters}}
\tablehead{
	\colhead{Cluster ID} & 
     \colhead{Short ID} &
	\colhead{RA} & 
	\colhead{Dec} & 
	\colhead{Spectroscopic} &
	\colhead{$N_{\rm{spec}}$} &
	\colhead{M$_{200}$} &
	\colhead{Additional} \\
	\colhead{} & 
	\colhead{} &
	\colhead{(J2000)} &
	\colhead{(J2000)} &
	\colhead{Redshift} &
	\colhead{} &
	\colhead{[$10^{14}\, \Msun$]} &
	\colhead{References} }
\startdata
ISCS J1432.4+3332\tablenotemark{a} & ID1 & 14:32:29.18 & 33:32:36.0 & 1.113 & 26 & $4.9^{+1.6}_{-1.2}$\tablenotemark{b} &  1, 2, 3, 4  \\
ISCS J1434.5+3427\tablenotemark{a} & ID2 & 14:34:30.44 & 34:27:12.3 & 1.238 & 19 & $2.5^{+2.2}_{-1.1}$\tablenotemark{b} & 1, 3, 4, 5  \\
ISCS J1429.3+3437\tablenotemark{a} & ID3 & 14:29:18.51 & 34:37:25.8 & 1.262 & 18 & $5.4^{+2.4}_{-1.6}$\tablenotemark{b} & 2, 3, 4  \\
ISCS J1432.6+3436\tablenotemark{a} & ID4 & 14:32:38.38 & 34:36:49.0 & 1.350 & 12 & $5.3^{+2.6}_{-1.7}$\tablenotemark{b} & 2, 3, 4  \\
ISCS J1434.7+3519\tablenotemark{a} & ID5 & 14:34:46.33 & 35:19:33.5 & 1.374 & 10 & $2.8^{+2.9}_{-1.4}$\tablenotemark{b} & 2, 3, 4  \\
ISCS J1432.3+3253\tablenotemark{a} & ID6 & 14:32:18.31 & 32:53:07.8 & 1.396 & 10 & $\ldots$ & 3, 4 \\
ISCS J1425.3+3250\tablenotemark{a} & ID7 & 14:25:18.50 & 32:50:40.5 & 1.400 & 7 & $\ldots$ & 3, 4 \\
ISCS J1438.1+3414\tablenotemark{a} & ID8 & 14:38:08.71 & 34:14:19.2 & 1.413 & 16 & $2.2^{+0.7}_{-0.6}$\tablenotemark{c} & 2, 3, 4, 6, 7 \\
ISCS J1431.1+3459\tablenotemark{a} & ID9 & 14:31:08.06 & 34:59:43.3 & 1.463 & 6 & $\ldots$ & 3, 4  \\
ISCS J1432.4+3250\tablenotemark{a} & ID10 & 14:32:24.16 & 32:50:03.7 & 1.487 & 11 & $2.5^{+1.5}_{-0.9}$\tablenotemark{c} & 3, 4, 7 \\
ISCS J1426.5+3508 & ID11 & 14:26:32.95 & 35:08:23.6 & 1.75 & 7 & $4.1\pm1.1$\tablenotemark{d} & 8, 9, 10, 11 
\enddata
\tablecomments{$^1$\citet{els06}; $^2$\citet{eis08}; $^3$\citet{bro13}; $^4$\citet{zei13}; $^5$\citet{bro06}; $^6$\citet{sta05}; $^7$\citet{bro11}; $^8$\citet{bro12}; $^9$\citet{gon12}; $^{10}$\citet{sta12}; $^{11}$\citet{bro15}}
\tablenotetext{a}{Cluster has H$\alpha$ measurements from {\it HST} grism spectroscopy (Section~\ref{sec:spectroscopy}) and targeted, deep MIPS imaging (Section~\ref{sec:mips}).}
\tablenotetext{b}{Weak lensing mass measurement from \citet{jee11}.}
\tablenotetext{c}{X-ray mass measurement from \citet{bro11}.}
\tablenotetext{d}{SZ mass measurement from \citet{bro12}.}

\end{deluxetable*}

\subsection{Spectroscopic Redshifts}
\label{sec:spectroscopy}

  Targeted follow-up campaigns by our group have obtained spectroscopic redshifts for galaxies and AGN in $z>1$ clusters using multi-object Keck optical spectroscopy and  Wide Field Camera 3 (WFC3) slitless NIR grism spectroscopy from {\it HST} \citep{kim08}.  The reader is directed to \citet{bro13},  \citet{zei13}, and reference therein for a detailed description of the targeted spectroscopy.   Some spectroscopic redshifts are additionally provided by the AGN and Galaxy Evolution Survey \citep[AGES;][]{koc12}, which includes spectroscopic redshifts for galaxies at $z\lesssim0.6$ and AGN at $z\lesssim3$.   Spectroscopic confirmation of a cluster is based on detection of at least five galaxies with spectroscopic redshifts in the range $\pm2000(1+\langle z_{spec}\rangle)$ km s$^{-1}$ and within a cluster-centric radius of 2 Mpc.  The number of spectroscopic redshifts in the main cluster sample for this work can be seen in Table~\ref{tbl:clusters}.

\subsection{New {\it Herschel}/PACS Imaging}
\label{sec:pacs}

We present new targeted imaging from the {\it Herschel Space Observatory} \citep{pil10} Photodetector Array Camera and Spectrometer \citep[PACS;][]{pog10} at 100 and 160$\,\mu$m, obtained during Open Time 2 observing (PID: OT2\_apope\_3).  Eleven clusters were observed  from  $1<z<1.8$ with integration times ranging from 270 $-$ 5040 s over 2-4 pointings in order to reach the same L$_{\rm IR}$ limit at all redshifts.  Each resulting map is centered on an individual cluster, with the exceptions of ID6 and ID10, which were observed in one map due to their small angular separation ($\sim$4 arcmin).   Each map covers a FOV of 7$^{\prime}$x7$^{\prime}$, a physical size of 2-3 Mpc in radius.   The central 5$^{\prime}$x5$^{\prime}$ of this area is uniform in depth, with a small loss in sensitivity toward the edges of the map (see Appendix~\ref{appendix:a}). 

Data reduction is performed using Unimap v5.4.0 \citep{tra11, pia12, pia15}, a generalized least-squares \citep[GLS;][]{lup93} mapmaker.  The individual astronomical observation requests (AORs) are first processed up to Level 1 in HIPE v10 \citep{ott10} and then converted to a Unimap usable format using UniHIPE.  Next, pre-processing, which removes  baseline drifts, offsets, jumps, and spikes due to cosmic rays, is applied, followed by the GLS mapmaker.  Finally, astrometry is corrected for each map by stacking point sources from a 5$\sigma$ MIPS 24$\,\mu$m catalog and removing any offsets in the stack.  The final PACS maps are in Jy pix$^{-1}$ with 1$^{\prime\prime}$ and 2$^{\prime\prime}$ pixel sizes at 100 and 160$\,\mu$m, respectively.  The rms sensitivity ranges from $\sim0.5-2\,$mJy in the central  5$^{\prime}$x5$^{\prime}$ region of each map (see Table~\ref{tbl:a} in Appendix~\ref{appendix:a}).

Given the resolution of PACS ($\rm{FWHM}\sim6.7^{\prime\prime}$ at 100$\,\mu$m and 11$^{\prime\prime}$ at 160$\,\mu$m) we expect the majority of sources and all cluster galaxies in our maps to be point sources.  PACS 100$\,\mu$m flux densities are extracted using PSF fitting at the known positions of IRAC sources in the SDWFS 5$\sigma$ 4.5$\,\mu$m catalog.  We note that though it is more common in the literature to use MIPS 24$\,\mu$m sources as  priors for {\it Herschel} source extraction \citep[e.g.][]{mag13}, deep MIPS imaging is not available for ID11.  In addition, using IRAC priors allow us to create more complete source catalogs as some PACS sources may not be detected by MIPS.  Within the redshift range relevant to this work, these MIPS dropouts will occur preferentially at $z\sim1.3$ due to silicate absorption \citep[e.g.][]{mag11}, making it more challenging to interpret changes in cluster populations around this epoch.  Given the depth of the SDWFS catalog, there is typically one IRAC source per PACS 100$\,\mu$m beam and we use visual inspection to identify the rare cases of blending. Blended sources are rejected from our samples.  We note that this excludes close merger systems, which likely contribute to cluster star formation activity. The local background is estimated in postage stamps around each IRAC positional prior and flux density uncertainties are measured from the full residual maps.  We test the robustness of our catalogs using Monte Carlo simulations and from these determine that we can measure accurate flux densities using priors down to the $2\sigma$ level.  We construct a 2$\sigma$ 100$\,\mu$m point source catalog and use these catalog positions as priors to extract the flux densities of sources in the PACS 160$\,\mu$m maps following the same PSF fitting procedure.  Based on simulations, our 100$\,\mu$m catalog is 70$\%$ complete at 5.5-1.3 mJy at $z=1-1.75$, which corresponds to L$_{\rm IR}\sim7\e{11}\,\Lsun$ over the redshift range probed.  For more details about the observations, source extraction, and completeness simulations, see Appendix~\ref{appendix:a}.

\subsection{Complementary Multi-Wavelength Photometry}

The Bo\"{o}tes field contains a wealth of multi-wavelength observations, with photometry from the X-ray to the radio.  The reader is referred to \citet{chu14} for a full description of the UV-to-MIR photometry used to derive the photometric redshifts used in this work (Section~\ref{sec:photoz}).  In the following, we describe the ancillary MIR-FIR photometry, as well as the X-ray observations used.

\subsubsection{{\it Spitzer}/IRAC, {\it Spitzer}/MIPS, and {\it Spitzer}/IRS Imaging}
\label{sec:mips}

 The IRAC Shallow Survey was followed up with three more observations as part of SDWFS \citep[][]{ash09}, providing a factor-of-two deeper IRAC catalog, with an aperture-corrected 5$\sigma$ limit of 5.2$\,\mu$Jy at 4.5$\,\mu$m ([4.5] = 18.83 mag).     {\it Spitzer}/MIPS observations are available from the MIPS AGN and Galaxy Evolution Survey \citep[MAGES;][]{jan10} to a 3$\sigma$ depth of 0.122 mJy at 24$\,\mu$m.  In addition, ten of the clusters in this work were targeted for deep MIPS 24$\,\mu$m observations, with 3$\sigma$ depths of 156 $\,\mu$Jy at $z=1$ to 36 $\,\mu$Jy at $z=1.5$, providing uniform depth in L$_{\rm IR}$.  For a complete description of the data reduction of the targeted MIPS observations and an analysis of MIPS-derived star formation properties of cluster galaxies, see \citet{bro13}. {\it Spitzer}/IRS peak-up imaging at 16$\,\mu$m was obtained for nine of the clusters in this work (GTO proposal $\#$50050, PI Fazio).  This imaging reaches a $5\sigma$ depth of 70$\,\mu$Jy (L$_{\rm IR}\sim3\e{11}\,\Lsun$ at $z\sim1.3$) over a 3.33$^{\prime}$x3.75$^{\prime}$ area centered on the deep MIPS pointings.  Photometry was extracted for point sources in the IRS image using IRAC positional priors.

\subsubsection{{\it Herschel}/SPIRE Imaging}
\label{sec:pacsphot}

{\it Herschel} Spectral and Photometric Imaging Receiver \citep[SPIRE;][]{gri10} observations at 250, 350, and 500$\,\mu$m are available in Bo\"{o}tes from the {\it Herschel} Multi-tiered Extragalactic Survey \citep[HerMES;][]{oli12}.  The SPIRE imaging reaches a 5$\sigma$ depth of 14 mJy at 250$\,\mu$m in the inner two square degrees of the Bo\"{o}tes field and 26 mJy over the remaining area, for a total of $\sim8$ square degrees of coverage. For a detailed description of the Bo\"{o}tes SPIRE imaging and our reduction of the data, see \citet{alb14}. 

\subsubsection{$Chandra$ X-ray Imaging}
\label{sec:xray}

Targeted X-ray observations of ten of the clusters in this study were obtained as a Cycle 10 {\it Chandra} program to a uniform exposure time of 40 ks.  In addition to identifying bright AGN, these X-ray observations were used to measure the X-ray emission of the intracluster medium (ICM), from which cluster halo masses can be derived.   For a full description of the X-ray data reduction and ICM measurements, the reader is directed to \citet[][]{bro11} \citep[see also][]{and11}.  The eleventh cluster, ID11, was observed as part of the XBo\"{o}tes Survey \citep{mur05, ken05} with an exposure time of 9.5 ks \citep[][]{sta12}.  Deeper 100 ks Chandra X-ray observations have recently been obtained for this cluster \citep{bro15}, yielding a halo mass estimate in good agreement with previous measurements.  XBo\"{o}tes is available across the entire Bo\"{o}tes field with exposure times of 5-15 ks, sufficient to detect unobscured moderate to luminous AGN \citep{ran03, mar13}.


\subsection{Photometric Redshifts}
\label{sec:photoz}

In this work, we adopt the new photometric redshift catalog of \citet{chu14}.  This catalog incorporates near-infrared observations from the Infrared Bo\"{o}tes Imaging Survey \citep[IBIS; ][]{gon10} with previously available UV-MIR photometry, providing greater accuracy for photometric redshifts at $z>1.5$.    Using up to 17 photometric bands, the photometric redshifts are calculated through SED fitting, using non-negative linear combinations of empirically derived templates \citep[for complete details, see][]{assef10}. An $R$-band luminosity prior from the Las Campanas Redshift Survey \citep{lin96} was used to avoid unphysical fits. The templates include a characteristic elliptical, spiral, and irregular (starburst), as well as an AGN template, which is introduced with a variable amount of internal reddening.  Each source was first fit with only galaxy templates.  Then they were fit by galaxy+AGN templates and an F-test was used to see if the addition of an AGN component significantly improved the goodness-of-fit \citep[see][for a detailed discussion]{chu14}.  Stellar and brown dwarf templates were also fit in order to identify Galactic sources.

Aside from the improved photometric redshift accuracy at $z>1.5$, arising from the inclusion of the IBIS NIR photometry, the advantage of this catalog over the \citet{bro06} photometric redshift catalog, which has been used in previous ISCS/IDCS and other Bo\"{o}tes field analyses, is that the SED fitting procedure provides a measure of the AGN content in each galaxy, described in Section~\ref{sec:fgal}.  We have tested that the \citet{chu14} and \citet{bro06} photometric redshift catalogs are consistent within their stated errors for galaxies out to $z=1.5$.  For the purposes of this work, we limit our photometric redshift catalog to sources with 4.5$\,\mu$m fluxes greater than 5.2$\,\mu$Jy (5$\sigma$). After removing stars and brown dwarfs, this catalog contains 281,779 sources.  

\citet{chu14} reports photometric redshift uncertainties of $\sigma/(1+z)$ = 0.040 for galaxies and $\sigma/(1+z)$ = 0.169 for AGN, with 5$\%$ outlier rejection.  As their analysis only includes sources that were unambiguously galaxies or unambiguously AGN as determined during the SED fitting, we further test the photometric redshift uncertainties in two ways.  First, we do a comparison with available spectroscopic redshifts for all galaxies and AGN, including ``composite'' objects that have a significant contribution from both.  We find $\sigma/(1+z)$ = 0.040 for sources dominated by host galaxy emission and $\sigma/(1+z)$ = 0.214 for sources dominated by AGN emission when these composite galaxies are included (see Section~\ref{sec:fgal} and Appendix~\ref{appendix:b}).  Second, since few spectroscopic redshifts are available for galaxies at $z>1.5$, we use pair statistics \citep{qua10, hua13} to test the photometric redshift uncertainties at high redshift.  Using this technique, we measure $\sigma/(1+z)$ = 0.054 for galaxies, with no significant dependence on redshift up to $z\sim2$.  For further details on these tests, see Appendix~\ref{appendix:b}.

\subsubsection{The Contribution from AGN: $F_{gal}$}
\label{sec:fgal}

\begin{figure*}[!ht]
\includegraphics[angle=270, scale=0.65]{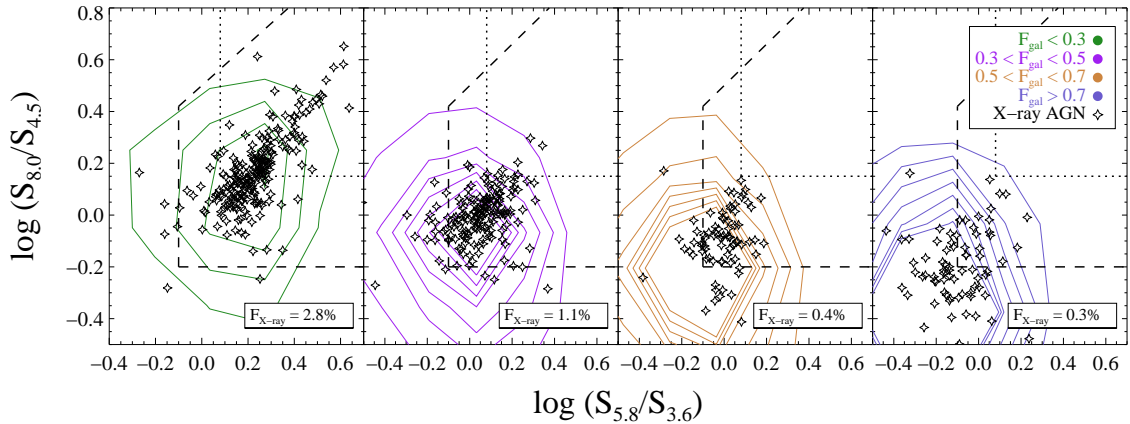}
\caption{\footnotesize{The IRAC colors of galaxies in the photometric redshift catalog broken into four subsets by F$_{\rm{gal}}$=L$_{\rm{gal}}$/L$_{\rm{tot}}$, which measures the relative fraction of luminosity in the UV-MIR from host galaxy emission versus the total luminosity (galaxy+AGN).  The contours show the number density of each subset in IRAC color space. The dashed line shows the \citet{lac04} criteria for MIR AGN selection, while the solid line shows the more conservative AGN selection from \citet{kir13}.  As F$_{\rm{gal}}$ increases, sources move from the region of IRAC color space associated with AGN to the region associated with non-AGN sources.  Star symbols denote X-ray AGN, which can be seen in all regions of IRAC color space.  The fraction of X-ray AGN in each F$_{\rm{gal}}$ subset, F$_{\rm{X-ray}}$, decreases with increasing F$_{\rm{gal}}$.}}
\label{fig:fgal}
\end{figure*}

As described in \citet{assef10} and \citet{chu14}, the AGN content of a source can be determined through SED fitting by measuring the contributions of the best-fit galaxy and AGN templates to the UV-MIR SED.  Specifically, we quantify the ratio of the UV-MIR luminosity from best-fit galaxy templates to the total UV-MIR luminosity (galaxy+AGN templates): F$_{\rm{gal}}\equiv\,$L$_{\rm{gal}}$/L$_{\rm{total}}$, with F$_{\rm{gal}}$=0.5 providing a useful dividing line between sources whose  luminosity is primarily from an AGN (F$_{\rm{gal}}<0.5$) versus those whose luminosity is primarily provided by the (host) galaxy (F$_{\rm{gal}}>0.5$).   This technique takes advantage of a broad wavelength range, providing a more complete selection, in terms of AGN type and the balance between AGN and host galaxy emission, than AGN indicators that use only limited wavelength windows or colors \citep[e.g.][]{hic09, men13, chu14}.   Checking against spectroscopic redshifts, \citet{assef10} found that the ability of this SED fitting technique to measure F$_{\rm{gal}}$ is relatively independent of the derived photometric redshift; even galaxies with large uncertainties in their photometric redshift have a small uncertainty in F$_{\rm{gal}}$.  

A detailed analysis of how this technique compares to other common AGN indicators, such as X-ray emission, spectral features, and MIR colors \citep[e.g.][]{lac04, ste05} was presented in \citet{chu14}.  We briefly summarize their findings here.  X-ray selected AGN were found to have a large range of F$_{\rm{gal}}$, with a tighter correlation for sources that are optically compact.  This is consistent with the large scatter in the MIR colors of X-ray AGN, which can lie outside of MIR AGN color space, and is likely due to soft X-ray observations being sensitive to lower luminosity AGN \citep{gor08, eck10, car08, men13}.  AGN at $z>1$ identified through optical spectral features, on the other hand, show a strong correlation with F$_{\rm{gal}}$, with $\sim80\%$ of these AGN having F$_{\rm{gal}}<0.5$.

Luminous AGN will have MIR SEDs that resemble a power-law and thus occupy a particular region of MIR color space  \citep[e.g.][]{lac04, ste05, don12, kir13}.  \citet{chu14} looked at unambiguous AGN, as identified through SED fitting, in MIR color space, finding that 57$\%$ and 75$\%$ were recovered by the \citet{lac04} and \citet{ste05} AGN selections respectively and 32$\%$ by the more conservative \citet{don12} selection.  It has been shown that, particularly for deeper MIR observations, the more conservative selection is necessary to remove dusty SFG interlopers from AGN samples selected by IRAC colors \citep{don12, kir13}.  This is consistent with our expectation that the SED fitting identifies lower luminosity AGN and composite objects with a significant host contribution in the MIR.

In Figure~\ref{fig:fgal}, we show the MIR colors of all galaxies in our photometric redshift catalogs in the relevant redshift range, $1<z<1.8$.  We break the galaxies into four categories:  F$_{\rm{gal}}<0.3$ (``AGN-dominated"), $0.3<\;$F$_{\rm{gal}}<0.5$ (``AGN-composite"), $0.5<\;$F$_{\rm{gal}}<0.7$ (``host-composite"), and F$_{\rm{gal}}>0.7$ (``host-dominated").   In general, sources trend from the region of MIR color space traditionally associated with AGN to that of star forming galaxies as a function of increasing F$_{\rm{gal}}$, with significant scatter and a significant fraction of AGN-dominated and AGN-composite source that would not be found in MIR selections alone. We note that X-ray AGN are found throughout MIR color space, with the X-ray AGN fraction decreasing with increasing  F$_{\rm{gal}}$.  The number of AGN detected in the X-ray is small for our dataset as the X-ray imaging available across the Bo\"{o}tes field is very shallow (see Section~\ref{sec:xray}).

\subsection{Stellar Masses}
\label{sec:masses}

Stellar mass estimates are available for sources in the SDWFS IRAC catalog \citep[see][]{bro13}, derived with optical and MIR photometry using iSEDfit \citep{mou13}, a Bayesian SED fitting code, with \citet{bru03} population synthesis models and a \citet{cha03} initial mass function (IMF).  Though individual mass errors are typically reported by iSEDfit to be $<0.2$ dex, a mass error of 0.3 dex is adopted in this work for all stellar mass estimates  in order to account for systematic uncertainties.  At $z>1$, this stellar mass catalog is 80$\%$ complete above log (M$_{\star}/\Msun) = 10.1$ \citep[see Figure 3 in][]{bro13}.  We note that these stellar mass estimates were derived assuming the photometric redshifts calculated in \citet{bro06}, which introduces an additional uncertainty.   This uncertainty, however, is small compared to the overall systematic uncertainties.  
\subsection{Cluster Membership}
\label{sec:membership}

Sources with spectroscopic redshifts are considered cluster members if they meet the criteria set in \citet{eis08}:  a spectroscopic redshift within 2000 km s$^{-1}$ of the systemic cluster velocity and within 2 Mpc of the cluster center.  We only consider robust spectroscopic redshifts when considering membership.  For sources with photometric redshifts, cluster membership is determined based on constraining the integral of the normalized probability distribution function (PDF).     For sources with F$_{\rm{gal}} > 0.5$, we integrate over the photometric redshift uncertainties calculated using pair statistics, $\sigma$ = $0.054(1+z_{\rm{cl}})$ (see Section~\ref{sec:photoz} and Appendix~\ref{appendix:b}).  Photometric redshift cluster members are thus identified as sources within 2 Mpc of the cluster center which satisfy the following criterion:

\begin{equation}
\label{eqn:membership}
\int_{z_{cl}-0.054(1+z_{cl})}^{z_{cl}+0.054(1+z_{cl})} P(z)\,dz \geq 0.3
\end{equation}
where $P(z)$ is the PDF of the photometric redshift.

Previous studies have found that the AGN fraction in galaxy clusters increases by two orders of magnitude from $z=0$ to $z=1$ \citep{gal09, mar13}. In order to account for the contribution to cluster star formation from the AGN population, we identify AGN as galaxies with F$_{\rm{gal}}<0.5$.  To determine cluster membership for these sources, we again adopt the photometric redshift uncertainty $\sigma$ = $0.054(1+z_{\rm{cl}})$ in Equation~\ref{eqn:membership}, rather than $\sigma$ = $0.214(1+z_{\rm{cl}})$ as was measured for F$_{\rm{gal}} < 0.5$ sources using pair statistics (see Appendix~\ref{appendix:b}), which would produce a sample strongly contaminated by field AGN.  This conservative approach gives us a better census of the total SF and AGN activity of cluster galaxies with minimal contamination; however, we note that our F$_{\rm{gal}}<0.5$ cluster member sample likely suffers from incompleteness and bias toward composite AGN, as it is more difficult to measure photometric redshifts for SEDs completely dominated by AGN power law emission.

Clusters ID6 ($z=1.396$) and ID10 ($z=1.487$) have an angular separation of only 4 arcmin ($\sim2\,$Mpc) between their centers.  Given the photometric redshift uncertainties, $\sim30\%$ of potential clusters members in the overlapping regions satisfy the cluster membership criteria for both clusters.  In order to avoid double-counting, we assign galaxies to the cluster for which they have the highest integrated photometric redshift PDF at the redshift of that cluster.  

Finally, the spectroscopic and photometric cluster member lists are checked for overlap.  Roughly 60$\%$ of the spectroscopic redshift cluster members have a match in the photometric redshift catalog and therefore a measurement of F$_{\rm{gal}}$. The remaining 40$\%$ are largely faint galaxies observed with the HST/WFC3 grism, which obtained a depth $\sim10$ times fainter than the SDWFS IRAC catalog at 4.5$\,\mu$m. The total number of cluster members identified is 569, with 142 spectroscopic redshift members and 328 (99) photometric redshift members with F$_{\rm{gal}}>0.5$ (F$_{\rm{gal}}<0.5$).  Roughly 10$\%$ of spectroscopic redshift members do not have a stellar mass estimate and can not be included in analyses that require a stellar mass or where a stellar mass cut is applied. 

\subsection{Matching Multi-Wavelength Catalogs}
\label{sec:matching}

Spectroscopic redshift cluster members are matched to the SDWFS IRAC catalog (search radius $r_s = 1^{\prime\prime}$) to determine stellar masses, IRAC, and PACS counterparts.  The spectroscopic cluster members are also checked for a counterpart in the photometric redshift catalog; if found, then UV-MIR is available through matched photometry catalogs \citep[see][for more details]{chu14}.  IRS 16$\,\mu$m and MIPS 24$\,\mu$m counterparts are searched for in the IRS and deep MIPS catalogs, using $r_s = 1^{\prime\prime}$ as the source extraction is based on IRAC priors \citep{bro13}.  If a MIPS detection is not available from the deep imaging because of incompleteness or being outside the field-of-view of the deep MIPS images, then the MAGES catalog is searched for a $>3\sigma$ detection within 3$^{\prime\prime}$ of the IRAC position.  

Photometric redshift cluster members automatically have matches to the full UV-MIR matched photometric catalogs used by \citet{chu14} and to the stellar mass catalog.  MIPS counterparts are determined as described above and PACS counterparts come directly from the IRAC priors.  X-ray detections were identified using a variable search radius to account for the off-axis PSF degradation.  All cluster members are visually inspected for blending with nearby bright sources in the PACS 100$\,\mu$m maps and removed from the cluster catalogs if blended.  Table~\ref{tbl:pacsstats} contains statistics for cluster members with PACS 100$\,\mu$m detections and Table~\ref{tbl:phot} contains the photometry for these members.

\begin{deluxetable*}{lccc}
\tabletypesize{\footnotesize}
\tablecolumns{4}
\tablewidth{0pt}
\tablecaption{Characteristics of PACS 100$\,\mu$m Detected Cluster Members\label{tbl:pacsstats}}
\tablehead{
	\colhead{} &
     \colhead{Number} & 
     \colhead{L$_{\rm IR}$ Range (Median) } &
	\colhead{SFR Range (Median) } \\
	\colhead{} &
	\colhead{Detected at 100$\,\mu$m\tablenotemark{a}} & 
	\colhead{$[10^{11} \Lsun]$} &
	\colhead{[$\Msun$ yr$^{-1}$]} }
\startdata
Spectroscopic Redshift  Members\tablenotemark{b} & 27 & 4-25 (7) & 50-360 (100) \\
Photometric Redshift Members (F$_{\rm gal}>0.5$) & 68 & 4-40 (8) & 60-590 (120)  \\
Photometric Redshift Members (F$_{\rm gal}<0.5$) & 34 & 4-30 (10) & 54-460 (170) 
\enddata
\tablenotetext{a}{Within 2 Mpc in projected radius.}
\tablenotetext{b}{Approximately $\sim40\%$ of spectroscopic redshift members do not have a match in the photometric redshift catalog and therefore do not have a measured F$_{\rm gal}$.} 
%\tablenotetext{c}{Number of spectroscopic cluster members within 1 Mpc of the cluster centers with log (M$_{\star}/\Msun)\geq10.1$ that are not AGN, as determined by F$_{\rm gal}$, X-ray, or visual inspection of the MIR.  See Section~\ref{sec:sed} and Figure~\ref{fig:speczsed}.}
\end{deluxetable*} 

\begin{turnpage}
\begin{deluxetable*}{lccccccccccc}
\tabletypesize{\footnotesize}
\tablecolumns{12}
\tablewidth{0pt}
\tablecaption{Photometry of PACS 100$\,\mu$m Detected Cluster Members\label{tbl:phot}\tablenotemark{a}}
\tablehead{
	\colhead{Cluster} &
	\colhead{Name} & 
     \colhead{RA} &
	\colhead{Dec} &
     \colhead{Redshift} &
     \colhead{Redshift} &
     \colhead{Cluster-centric} &
	\colhead{F$_{\rm 4.5\mu m}$} &
	\colhead{F$_{\rm 16\mu m}$} &
	\colhead{F$_{\rm 24\mu m}$} &
	\colhead{F$_{\rm 100\mu m}$} &
	\colhead{F$_{\rm 160\mu m}$} \\
	\colhead{ID} &
	\colhead{} &
	\colhead{(J2000)} &
	\colhead{(J2000)} &
	\colhead{} &
	\colhead{Type} &
	\colhead{Radius [Mpc]} &
	\colhead{[mJy]} &
	\colhead{[mJy]} &
	\colhead{[mJy]} &
	\colhead{[mJy]} &
	\colhead{[mJy]}  
}
\startdata
ISCS J1432.4+3332 & J143230.1+332927 & 14:32:30.15 & 33:29:27.7  & 1.07 &  Photoz & 1.6 & 0.032$\pm$0.001 & $\ldots$ & 0.21$\pm$0.02 & 4.9$\pm$2.2 & 3$\pm$6 \\
ISCS J1432.4+3332 & J143217.2+332959 & 14:32:17.17 & 33:29:59.6 & 1.13 & Photoz & 1.8 & 0.030$\pm$0.001 & $\ldots$ & $\ldots$ & 7.8$\pm$2.1 & 15$\pm$5 \\
ISCS J1432.4+3332 & J143228.9+333040 & 14:32:28.91 & 33:30:40.3 & 1.111 & Specz & 1.0 & 0.030$\pm$0.001 & $\ldots$ & 0.20$\pm$0.02 & 3.8$\pm$1.8 & 7$\pm$4 \\
ISCS J1432.4+3332 & J143235.5+333054 & 14:32:35.47 & 33:30:54.6 & 1.02 & Photoz & 1.1 & 0.045$\pm$0.001 & 0.29$\pm$0.01 & 0.36$\pm$0.02 & 7.8$\pm$2.0 & 5$\pm$4 \\
ISCS J1432.4+3332 & J143228.4+333152 & 14:32:28.42 & 33:31:52.3 & 1.10 & Photoz & 0.4 & 0.030$\pm$0.001 & 0.09$\pm$0.01 & 0.06$\pm$0.01 & 4.8$\pm$2.1 & 5$\pm$5 \\
ISCS J1432.4+3332 & J143234.2+333239 & 14:32:34.25 & 33:32:39.9 & 1.098 & Specz & 0.5 & 0.030$\pm$0.001 & 0.02$\pm$0.01 & $\ldots$& 3.5$\pm$1.5 & $\ldots$ \\
ISCS J1432.4+3332 & J143227.4+333254 & 14:32:27.40 & 33:32:54.0 & 1.03 & Photoz & 0.2 & 0.069$\pm$0.001 & 0.32$\pm$0.01 & 0.57$\pm$0.03 & 12$\pm$2.0 & 8$\pm$4 \\
ISCS J1432.4+3332 & J143246.0+333258 & 14:32:46.04 & 33:32:58.1 & 1.03 & Photoz & 1.8 & 0.021$\pm$0.001 & $\ldots$ & 0.12$\pm$0.02 & 4.3$\pm$2.1 & 2$\pm$5 \\
ISCS J1432.4+3332 & J143242.4+333339 & 14:32:42.41 & 33:33:39.1 & 1.19 & Photoz & 1.5 & 0.038$\pm$0.001 & 0.47$\pm$0.03 & 0.42$\pm$0.02 & 9.7$\pm$1.9 & 23$\pm$8 \\
ISCS J1432.4+3332 & J143231.5+333344 & 14:32:31.53 & 33:33:44.2 & 1.18 & Photoz & 0.6 & 0.042$\pm$0.001 & 0.34$\pm$0.01 & 0.38$\pm$0.02 & 8.4$\pm$2.5 & 3$\pm$4 \\
\multicolumn{12}{c}{} \\
\multicolumn{12}{c}{$\ldots$} \\
\multicolumn{12}{c}{} \\
ISCS J1426.5+3508 & J142649.1+350948 & 14:26:49.09 & 35:09:48.2 & 1.84 & Photoz & 1.9 & 0.014$\pm$0.001 & $\ldots$ &  0.14$\pm$0.04 & 1.8$\pm$0.8 & 5$\pm$2 \\
ISCS J1426.5+3508 & J142620.2+351059 & 14:26:20.17 & 35:10:59.5 & 1.68 & Photoz & 1.9 & 0.026$\pm$0.001 & $\ldots$ & 0.49$\pm$0.05 & 7.6$\pm$0.7 & 8$\pm$3 \\
ISCS J1426.5+3508 & J142630.3+351103 & 14:26:30.26 & 35:11:03.5 & 1.62 & Photoz & 1.4 & 0.021$\pm$0.001 & $\ldots$ & 0.18$\pm$0.05 & 2.6$\pm$0.6 & 7$\pm$2
\enddata
\tablenotetext{a}{Table~\ref{tbl:phot} is published in its entirety in the electronic edition of ApJ. A portion is shown here for guidance regarding its form and content.}
\end{deluxetable*} 
\end{turnpage}

\section{Results}
\label{sec:analysis}

\subsection{Spectral Energy Distributions of {\it Herschel}\hyp selected Cluster Galaxies}
\label{sec:sed}

In order to examine the impact of environment on the shape of the spectral energy distribution of star-forming galaxies, we use our {\it Herschel}-selected sample to compare the average SED of cluster galaxies to empirical SED templates developed using field galaxies.  For this analysis, we select cluster galaxies within the virial radius ($r<1\,$Mpc) to minimize contamination from field galaxies, requiring a detection in at least the 4.5$\,\mu$m and 100$\,\mu$m bands, and a stellar mass of log (M$_{\star}/\Msun$) $\geq 10.1$.  We further break this sample into three subsamples by membership (spectroscopic or photometric redshift) and AGN contribution.

\begin{figure*}[!ht]
\centering
\includegraphics[angle=270, scale=0.55]{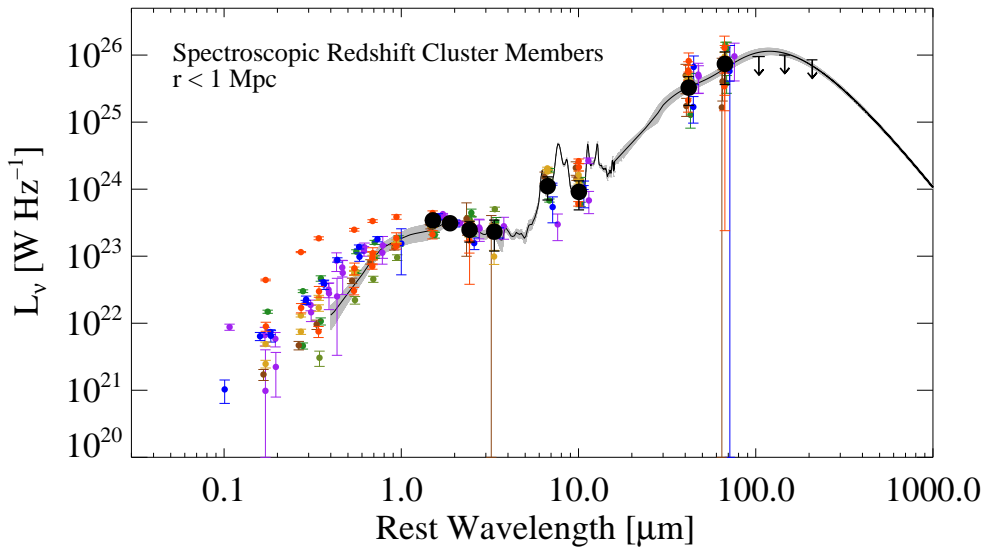}
\caption{\footnotesize{Optical-to-FIR SEDs of PACS-selected star-forming spectroscopic cluster members with log (M$_{\star}/\Msun)\geq10.1$ within $r<1\,$Mpc of the cluster cores, normalized at 4.5$\,\mu$m.  Small symbols show individual cluster members, while the large, black circles are the weighted average of all sources.  Photometry at rest wavelengths longward of 100$\,\mu$m were obtained by stacking on the SPIRE 250, 350, and 500$\,\mu$m images.  The SPIRE datapoints as shown are the (detected) stacked values; however, these values are formally upper limits (represented by the arrows) due to the fact that no correction has been applied for additional flux in the stacks from source confusion. AGN, selected via F$_{\rm{gal}}<0.5$, X-ray, or power law emission in the MIR as determined by visual inspection, are not included. The average SED of these cluster members is consistent with an empirically derived SED template for field galaxies at $z\sim1-2$ \citep[][]{kir12, kir15}, as shown by the black line, with template uncertainties denoted by the shaded region.}}
\label{fig:speczsed}
\end{figure*}

We start with our most robust cluster sample: spectroscopic redshift members.  Since there are only a few sources in this subsample with evidence for an AGN, we remove sources that have F$_{\rm{gal}}<0.5$, are X-ray detected, or have a MIR SED dominated by power-law emission as determined by visual inspection, leaving a purely star-forming sample of 15 spectroscopic cluster members that meet the criteria outlined above.  The optical to FIR SEDs of these galaxies can be seen in Figure~\ref{fig:speczsed}, normalized at rest-frame 4.5$\,\mu$m.    Overlaid in large, black circles are the average luminosities weighted by the inverse variance in the IRAC 3.6, 4.5, 5.6, and 8.0$\,\mu$m, IRS 16$\,\mu$m, MIPS 24$\,\mu$m, and PACS 100 and 160$\,\mu$m bands.  We additionally display the average luminosity in the SPIRE bands, which is determined through stacking following the procedure outlined in \citet{alb14}. Because of the small number of stacked objects, we do not attempt to correct for boosting in the SPIRE bands due to source confusion and clustering \citep[see][]{vie13, alb14}, so these points are formally upper limits even though they are detected in the stack. Cluster members in ID7 are not included in the stack due to being outside the SPIRE FOV.

We compare our cluster galaxy average photometry to a library of empirically derived SED templates developed for IR-luminous field galaxies at $z\sim1-2$ \citep[][]{kir12, kir15}.  This library is chosen for several reasons: 1) the field galaxies surveyed are IR-selected, with well sampled FIR SEDs from {\it Herschel} photometry, 2) they cover similar ranges in redshift ($z\sim1-2$) and expected  L$_{\rm{IR}}$  as our cluster galaxies, and 3) IRS spectroscopy was used to determine the AGN content of each galaxy through SED decomposition in the MIR \citep{pop08, kir12}.  The library contains templates for a range (0-100$\%$) of AGN content in the MIR SED, with measured contributions from the AGN to the total  L$_{\rm{IR}}$ for each template.

We perform the comparison using $\chi^2$ minimization between the average photometry (with bootstrapped errors) in the IRAC to PACS bands and each template in the \citet{kir15} library, each of which represents a SFG at high redshift with increasing AGN content.   Though we have optical photometry, the templates are not well optimized in the optical and so we do not include these bands in the comparison.  The SPIRE stacked photometry is also not included in the fits as we cannot accurately account for bias due to source confusion and clustering; instead, we check that our stacks, which are formally upper limits, are consistent with the best-fit templates. We find that for spectroscopic redshift cluster members, the average photometry is well described by a purely star forming field galaxy template (reduced $\chi^2\sim0.2$; Figure~\ref{fig:speczsed}).

\begin{figure*}[!ht]
\centering
\includegraphics[angle=270, width=0.8\linewidth]{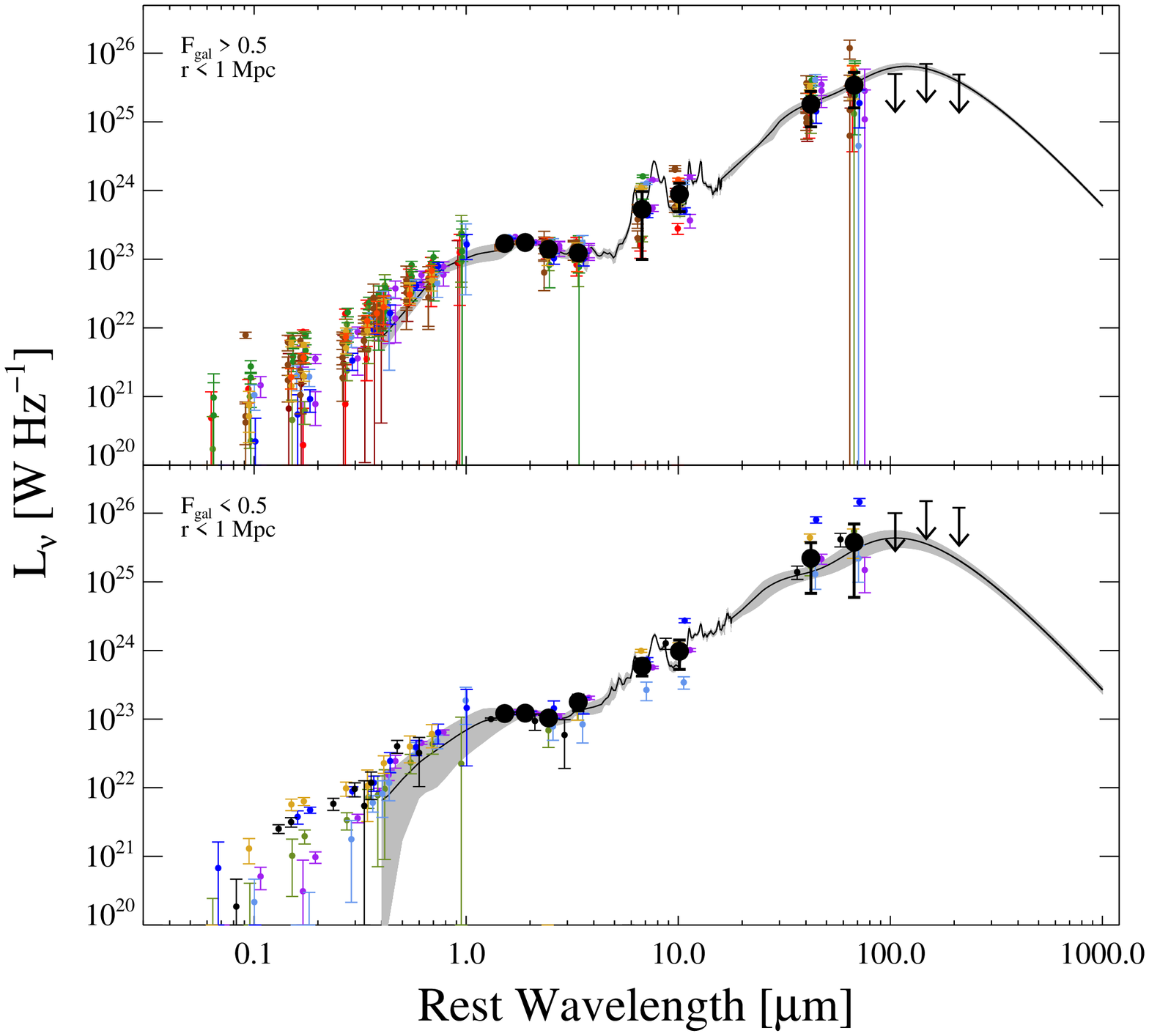}
\caption{As in Figure~\ref{fig:speczsed}, but for photometric redshift cluster members at $r<1\,$Mpc.  Small, colored symbols are individual galaxies, while the large, black circles show the weighted average of all galaxies.  Upper panel:  PACS-selected star forming cluster galaxies (F$_{\rm{gal}}>0.5$).  Lower panel:  PACS-selected AGN (F$_{\rm{gal}}<0.5$).The solid lines and shaded regions show representative SFG and AGN templates from \citet{kir15} in the upper and lower panels, respectively.  IR-luminous cluster galaxies at all radii have NIR-to-FIR SEDs that are consistent, on average, with field galaxy templates.}
\label{fig:photozsed}
\end{figure*}

We repeat this analysis for photometric redshift cluster members within the virial radius ($r\sim1\,$Mpc; Figure~\ref{fig:photozsed}) splitting the sources into subsamples with F$_{\rm gal}>0.5$ (19) and F$_{\rm gal}<0.5$ (6), where the latter consists of sources with $>50\%$ contribution from the AGN in the optical-MIR SED.  We find that the F$_{\rm gal}>0.5$ cluster galaxies are well fit (reduced $\chi^2\sim2$) by the purely star forming template, as was found for the spectroscopic redshift sample.  The F$_{\rm gal}<0.5$ sample is best fit (reduced $\chi^2\sim3$) by a template with $50\%$ of its MIR SED coming from an AGN.   This is consistent with our AGN selection (based on the optical-MIR SED) and our use of photometric redshifts for cluster membership, which will be biased against the most luminous AGN (see Section~\ref{sec:fgal}).

These results demonstrate that the near- to far-infrared SEDs ($\sim1-200\,\mu$m) of cluster galaxies show no significant deviation from the SEDs of field galaxies at similar redshifts, on average.  Additional detections on the Rayleigh-Jeans tail of the dust emission are required to directly quantify dust properties of cluster galaxies and probe the full FIR SED; submillimeter observations in ID11 at $z=1.75$ cluster will be presented in a future work (Alberts et al., in prep). 

We utilize the best fit templates to derive total infrared luminosities for each galaxy in each subsample,  by normalizing by the PACS 100$\,\mu$m flux and then integrating under the template to  measure L$_{\rm{IR}}\equiv\,$L[8-1000$\,\mu$m].  Though we do not have individual detections at longer wavelengths to show that the FIR SED is fully consistent, we find that the stacked submillimeter flux densities from SPIRE are consistent with the best-fit templates.  We also note that the total L$_{\rm{IR}}$ is dominated by the shorter wavelength FIR emission ($\sim70\%$ of the IR luminosity is accounted for at 8-100$\,\mu$m) where we have good coverage of the SED.    We calculate the SFR following the relation from \citet{mur11b}, 
\begin{equation}
\label{eqn:murphy}
\mathrm{SFR\, [\Msun\,yr^{-1}}] = 1.47\e{-10}\,\mathrm{ L^{SF}_{IR} }\,[\Lsun],
\end{equation}
where L$^{\rm{SF}}_{\rm{IR}}$ is the contribution to the L$_{\rm{IR}}$ coming from SF only.  For F$_{\rm gal}>0.5$ sources, L$^{\rm{SF}}_{\rm{IR}}$ is equal to the total L$_{\rm{IR}}$.  For  F$_{\rm gal}<0.5$ sources, SED decomposition of the best fit template was used to determine that 21$\%$ of the L$_{\rm{IR}}$ is due to the AGN and so we correct for this factor as L$^{\rm{SF}}_{\rm{IR}} = 0.79\times \rm{L}^{\rm{tot}}_{\rm{IR}}$ \citep[for more details, see][]{kir15}.  Table~\ref{tbl:properties} gives the stellar mass, F$_{\rm gal}$, L$^{\rm SF}_{\rm IR}$, and SFR for our PACS-detected cluster members.  Our L$_{\rm{IR}}$ to SFR conversion assumes a \citet{kro01} IMF, which has a similar normalization as the Chabrier IMF assumed for our stellar mass estimates \citep[see][]{spe14}.  

\begin{deluxetable*}{lcccccc}
\tabletypesize{\footnotesize}
\tablecolumns{7}
\tablewidth{0pt}
\tablecaption{Derived Properties of PACS 100$\,\mu$m Detected Cluster Members\label{tbl:properties}\tablenotemark{a}}
\tablehead{
	\colhead{Cluster} &
	\colhead{Name} &
	\colhead{Redshift} &  
     \colhead{$\log$ M$_{\star}$\tablenotemark{b}} &
	\colhead{F$_{\rm gal}$} &
     \colhead{L$^{\rm SF}_{\rm IR}$} &
     \colhead{SFR} \\
	\colhead{ID} &
	\colhead{} &
	\colhead{} &
	\colhead{[$\Msun$]} &
	\colhead{} &
	\colhead{[$10^{11}\,\Lsun$]} &
	\colhead{$\Msun$ yr$^{-1}$} 
}
\startdata
ISCS J1432.4+3332 & J143230.1+332927 & 1.07 & 10.8 & 0.95 & 7$\pm$3 & 100$\pm$40 \\
ISCS J1432.4+3332 & J143217.2+332959 & 1.13 & 10.9 & 0.59 & 10$\pm$3 & 150$\pm$40 \\
ISCS J1432.4+3332 & J143228.9+333040 & 1.111 & 10.6 & 0.87 & 5$\pm$2 & 80$\pm$40 \\
ISCS J1432.4+3332 & J143235.5+333054 & 1.02 & 10.9 & 0.78 & 10$\pm$3 & 150$\pm$40 \\
ISCS J1432.4+3332 & J143228.4+333152 & 1.10 & 10.7 & 0.78 & 6$\pm$3 & 90$\pm$40 \\
ISCS J1432.4+3332 & J143234.2+333239 & 1.098 & 10.7 & 1.00 & 5$\pm$2 & 70$\pm$30 \\
ISCS J1432.4+3332 & J143227.4+333254 & 1.03 & 11.2 & 0.47 & 15$\pm$3 & 220$\pm$40 \\
ISCS J1432.4+3332 & J143246.0+333258 & 1.03 & 10.6 & 0.74 & 6$\pm$3 & 80$\pm$40 \\
ISCS J1432.4+3332 & J143242.4+333339 & 1.19 & 10.9 & 0.45 & 12$\pm$3 & 180$\pm$40 \\
ISCS J1432.4+3332 & J143231.5+333344 & 1.18 & 11.1 & 0.52 & 11$\pm$3 & 160$\pm$50 \\
\multicolumn{7}{c}{} \\
\multicolumn{7}{c}{$\ldots$} \\
\multicolumn{7}{c}{} \\
ISCS J1426.5+3508 & J142649.1+350948 & 1.84 & 9.3 & 0.09 & 7$\pm$3 & 100$\pm$50 \\
ISCS J1426.5+3508 & J142620.2+351059 & 1.68 & 10.0 & 0.29 & 29$\pm$3 & 430$\pm$50 \\
ISCS J1426.5+3508 & J142630.3+351103 & 1.62 & 10.8 & 0.77 & 12$\pm$3 & 170$\pm$40
\enddata
\tablenotetext{a}{Table~\ref{tbl:properties} is published in its entirety in the electronic edition of ApJ. A portion is shown here for guidance regarding its form and content.}
\tablenotetext{b}{An error of 0.3 dex is adopted for all stellar mass measurements (see Section~\ref{sec:masses}).}
\end{deluxetable*} 

\subsection{Star Formation Properties of High Redshift Cluster Members}
\label{sec:sfprop}

\begin{figure*}[!ht]
% SFR and SSFR MS
\begin{minipage}[b]{0.45\linewidth}
\centering
\includegraphics[angle=270, trim=10mm 7mm 0 0, clip, scale=0.35]{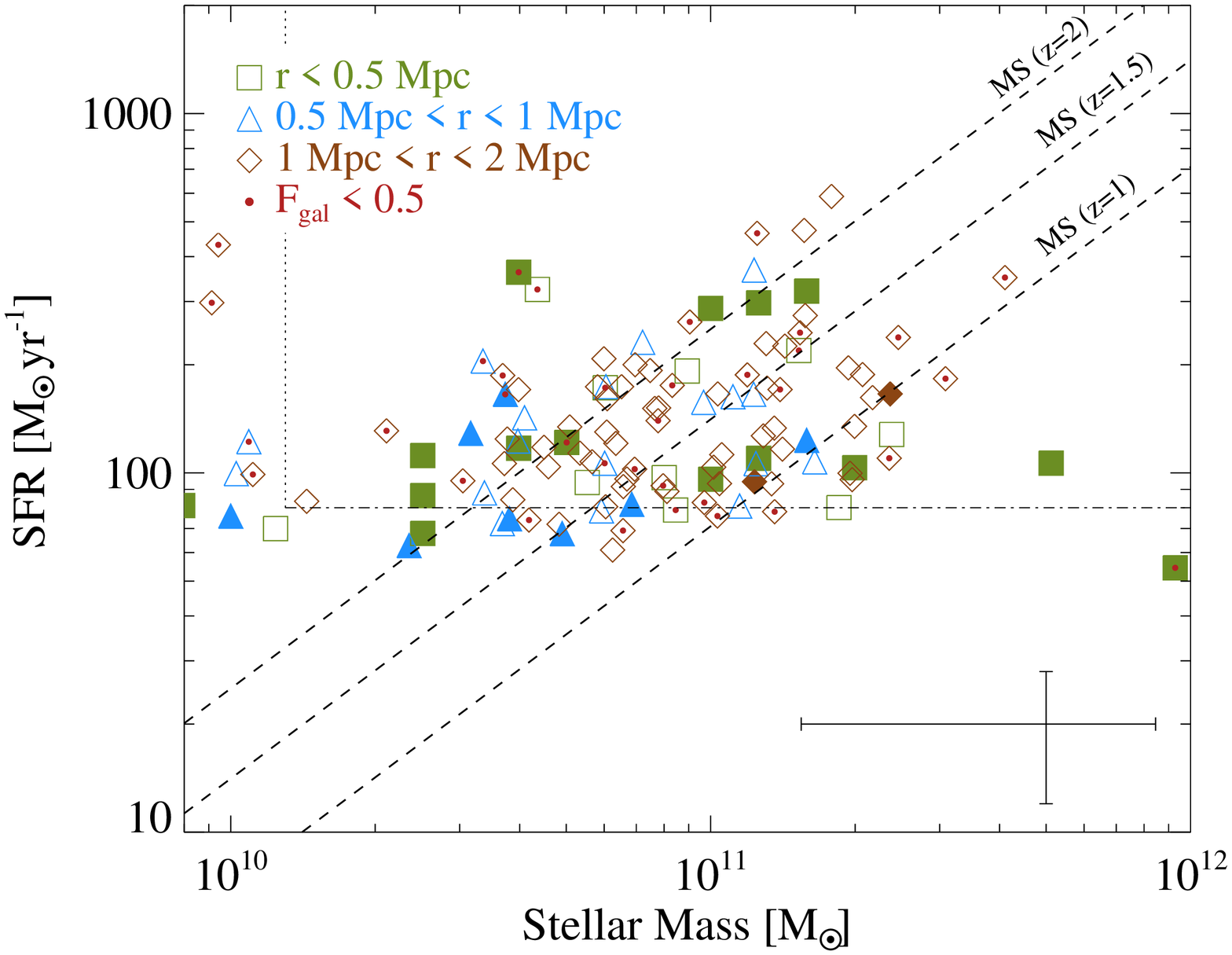}
\end{minipage}
\hspace{0.5cm}
\begin{minipage}[b]{0.45\linewidth}
\centering
\includegraphics[angle=270, trim=10mm 7mm 0 0, clip, scale=0.35]{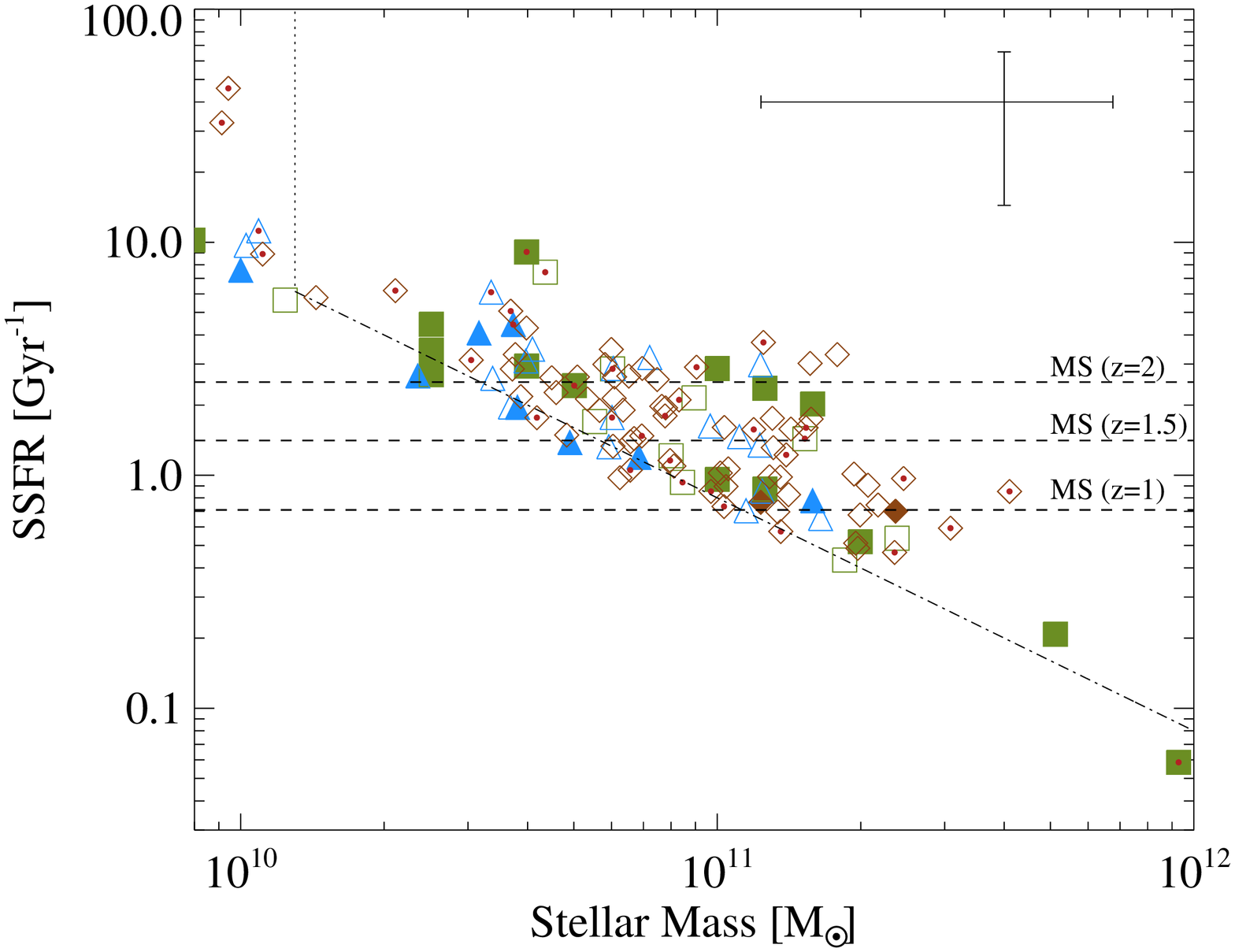}
\end{minipage}
\caption{\footnotesize{Left: The SFR of IR-luminous cluster galaxies as a function of stellar mass in three radial bins.  Right:  The SSFR of the same galaxies as a function of stellar mass.  Solid symbols indicate spectroscopic redshift members. AGN (F$_{\rm{gal}}<0.5$) are marked by small red dots.  The dot-dashed lines in both panels show the 50$\%$ completeness limit in SFR ($\sim80\,\Msun$ yr$^{-1}$) for SFGs.  The vertical dotted lines indicate the 80$\%$ mass completeness limit, log (M$_{\star}/\Msun) = 10.1$.  The dashed lines show the Main Sequence (MS) at $z=1$, 1.5, and 2 \citep{elb11}.  The large error bars show the systematic uncertainties on the SFR, SSFR, and M$_{\star}$.} }
\label{fig:pacsms}
\end{figure*}

Table~\ref{tbl:pacsstats} summarizes the L$_{\rm IR}$ and SFR characteristics of our {\it Herschel}-selected cluster members.  The distribution of SFRs and specific-SFRs (SSFR$\equiv$SFR/M$_{\star}$) as a function of stellar mass and radius can be seen in Figure~\ref{fig:pacsms}.   The dotted line denotes the stellar mass cutoff (log (M$_{\star}/\Msun) \geq 10.1$) which is adopted in the following analyses.  The dot-dash line indicates the 50$\%$ SFR completeness level for star-forming galaxies ($\sim\!80\,\Msun$ yr$^{-1}$), based on the PACS completeness and the SFG template.  This SFR completeness limit will be $\sim15\%$ lower for F$_{\rm{gal}}<0.5$ sources, marked by red dots, as their PACS flux (and thus L$_{\rm{IR}}$) includes a contribution from AGN emission.  This AGN contribution to the L$_{\rm{IR}}$ is removed to determine the SFRs seen in Figure~\ref{fig:pacsms}, as described in Section~\ref{sec:sed}.  For reference, the Main Sequence (MS) of galaxies is shown at $z=1$, $z=1.5$, and $z=2$ (dashed lines), adopted from \citet{elb11}, and corrected to a \citet{kro01} IMF and the \citet{mur11b} L$_{\rm{IR}}$ to SFR conversion:

\begin{equation}
\mbox{SSFR}_{\mathrm{MS}}\,[\mbox{Gyr}^{-1}] = 36.2 \times t_{\rm{cosmic}}^{-2.2}
\end{equation}
where $t_{\rm{cosmic}}$ is the cosmic time since the Big Bang.  It should be noted that our sample is SFR-limited and so does not probe the MS for the full range of M$_{\star}$ above our mass cutoff.  Assuming a scatter around the MS of a factor of two \citep{elb11} and our 50$\%$ completeness limit, we are unlikely to detect MS galaxies below log (M$_{\star}/\Msun) < 10.8$ [log (M$_{\star}/\Msun) < 10.5$] at $z=1$ [$z=1.5$].  

Figure~\ref{fig:pacsms} shows that our spectroscopically confirmed cluster members have a comparable range in SFR, M$_{\star}$, and SSFR to photometric redshift members.  Similarly, cluster members with significant AGN content (F$_{\rm gal}<0.5$), denoted by red dots, span a similar region as those without, though the former has a higher median SFR (Table~\ref{tbl:pacsstats}).  Our full {\it Herschel}-selected cluster galaxy sample falls in the general region that is described by the MS in field galaxy studies \citep[e.g.][]{elb11,rod10,rod14}, indicating that cluster galaxies have similar SFRs and SSFRs as field galaxies at this epoch.  However, as galaxies in cluster cores show significant quenching at $z\lesssim1$ \citep[e.g.][]{pat09, muz12, bro13}, with lower average SFRs relative to the field \citep{alb14}, there are clearly important environmental differences driving the evolution of cluster galaxies.  Our results are in good agreement with the analysis of MIPS-selected cluster galaxies drawn from an overlapping cluster sample in \citet{bro13}.

\subsubsection{Star Formation as a Function of Cluster-Centric Radius}
\label{sec:sfradius}

\begin{figure*}[!ht]
% Average SFR and SSFR, zbins
\centering
\includegraphics[angle=270, trim=0 5mm 0 0, clip, scale=0.55]{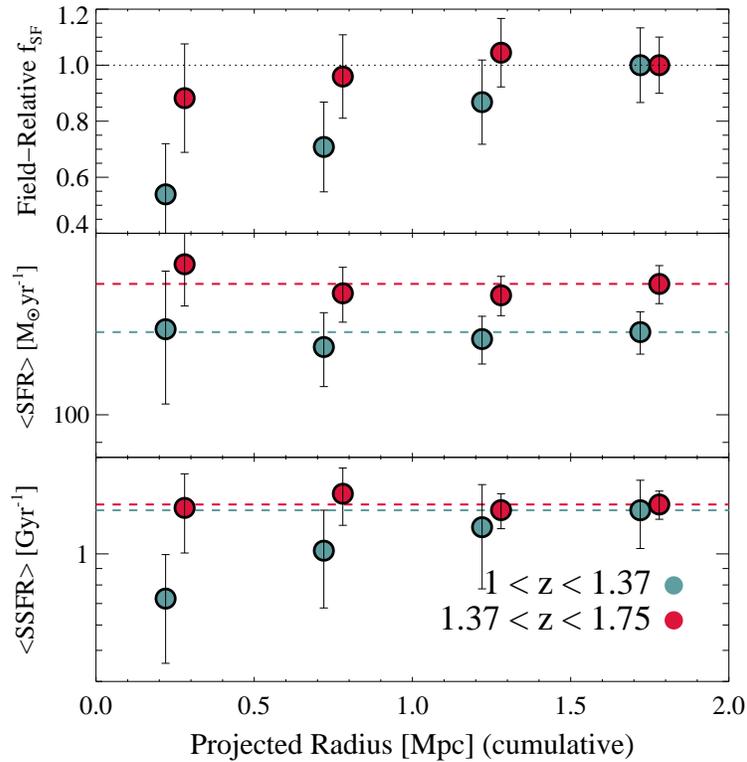}
\caption{\footnotesize{The fraction of PACS 100$\,\mu$m-detected galaxies (upper panel), average SFR (middle panel), and average SSFR (bottom panel) of IR-luminous cluster galaxies with $\log$ (M$_{\star}/\Msun) > 10.1$ as a function of projected radius.  Clusters were split into two redshift bins with median redshifts of 1.26 (blue) and 1.46 (red).   The star forming fraction, f$_{\rm SF}$, has been normalized to the outermost radial bin at $\sim2\times$ the virial radius, which is taken to be representative of the field.  The $\langle \rm SFR \rangle$ and $\langle \rm SSFR \rangle$ are not normalized; the dashed lines show the representative field values in each redshift bin.  This demonstrates the evolution of PACS-selected field galaxies with during this epoch. The average trends for the higher redshift ($1.37<z<1.75$, red) clusters are flat as a function of radius, consistent with little to no environmental quenching.  The lower redshift clusters (blue), conversely, show decreases in f$_{\rm{SF}}$ and $\langle \rm{SSFR} \rangle$ at the innermost radii, indicating quenching in excess of what is observed in the field.  Errors are binomial for the fraction (upper) and determined through bootstrapped resampling for the average quantities (middle, lower).  Bootstrapped errors encompass the spread in the population.  All quantities are cumulative with increasing radius.}}
\label{fig:sf1}
\end{figure*}

We next look at the fraction of PACS 100$\,\mu$m detected cluster members as a function of projected radius  (Figure~\ref{fig:sf1}, top). We split our cluster sample into two redshift bins: $1<z<1.37$ and $1.37<z<1.75$, allowing us to compare to the analysis of MIPS-selected cluster galaxies presented in \citet{bro13}. In order to highlight environmental trends, we make the assumption that our outermost radial bin is a good approximation of the field and normalize by this value.  We find that, moving from low to high redshift, the fraction of IR-luminous cluster galaxies flattens into the cluster cores, going from $\sim50\%$ of the field value at $z<1.37$ to consistent within 1$\sigma$ with the field value at $z>1.37$ in the innermost radial bin. Within the very centers of the clusters ($r<250\,$kpc), $\sim30\%$ of the cluster galaxies are PACS-detected at $z>1.37$, versus $\sim15\%$ in the lower redshift clusters.  Figure~\ref{fig:sf1} (top) demonstrates that: 1) IR-luminous cluster galaxies at $z\ga1.4$ are present in the cluster cores in numbers approaching that in the field, and 2) over a relatively short timescale ($\lesssim1$ Gyr), a significant fraction of these galaxies must be quenched below our detection limit, in excess of the normal evolution of field galaxies along the MS as a function of redshift.  We repeat this analysis for cluster members with log (M$_{\star}/\Msun) > 10.8$ to confirm that these trends are not driven by our sensitivity to the MS as a function of redshift.

We compare our results in Figure~\ref{fig:sf1} (top) to Figure 6 in \citet{bro13}, which shows a strong transition from $z=1$ to $z=1.5$ in the fraction of SFGs.  This includes a {\it rising} SFG fraction in the cluster cores ($r<0.5\,$Mpc), in excess of the field, at $z\gtrsim1.4$ for MIPS-selected galaxies with L$_{\rm IR}\gtrsim3\e{11}\,\Lsun$.  Our findings are in good agreement with these results as a function of redshift, with the {\it Herschel}-selected IR-luminous galaxy fraction (including galaxies with AGN, which were removed from the \citet{bro13} samples) flattening into the cluster cores at $z\gtrsim1.4$.  We do not see a comparable rise in the IR-luminous fraction above the field, however, which implies that environmentally driven processes resulting in the excess primarily boost galaxies up to moderate SFRs at this epoch.  Constraints on the IR-luminous fraction in clusters at higher redshift are necessary to disentangle whether such processes cannot produce bright IR galaxies or whether these  galaxies simply evolved earlier (i.e. downsizing). 

The middle and bottom panels of Figure~\ref{fig:sf1} show the average SFR and SSFR as functions of projected radius and redshift.  Errors are determined using bootstrapping and thus represent the spread in the SF properties of the full population.  The average SFR is consistent with being flat with projected radius for both redshift bins (Figure~\ref{fig:sf1}, middle), with cluster galaxies in the cluster cores having field-like SFRs. 

In the bottom panel, we see that  $\langle\,\rm{SSFR}\,\rangle$ follows a different trend with projected radius between the two redshift bins.  At lower redshift, we see a decline in the average SSFR by a factor of $\sim2$ into the cluster cores, indicating that these galaxies are not forming stars at the same rate, {\it for their mass}, as their counterparts in the field.  At higher redshift, we see no such environment effect in the SSFR, which is flat into the cluster cores, a trend previously observed in \citet{bro13}.  This result demonstrates that the IR-luminous cluster galaxies are undergoing the transition first proposed in \citet{bro13}, which roughly marks the epoch in which the quenching of cluster galaxies in the cluster cores becomes effective. 

\begin{figure}[!ht]
% Average SFR and SSFR, mbins
\centering
\makebox[\columnwidth]{\includegraphics[angle=270, , trim=10mm 7mm 0 0, clip, scale=0.45]{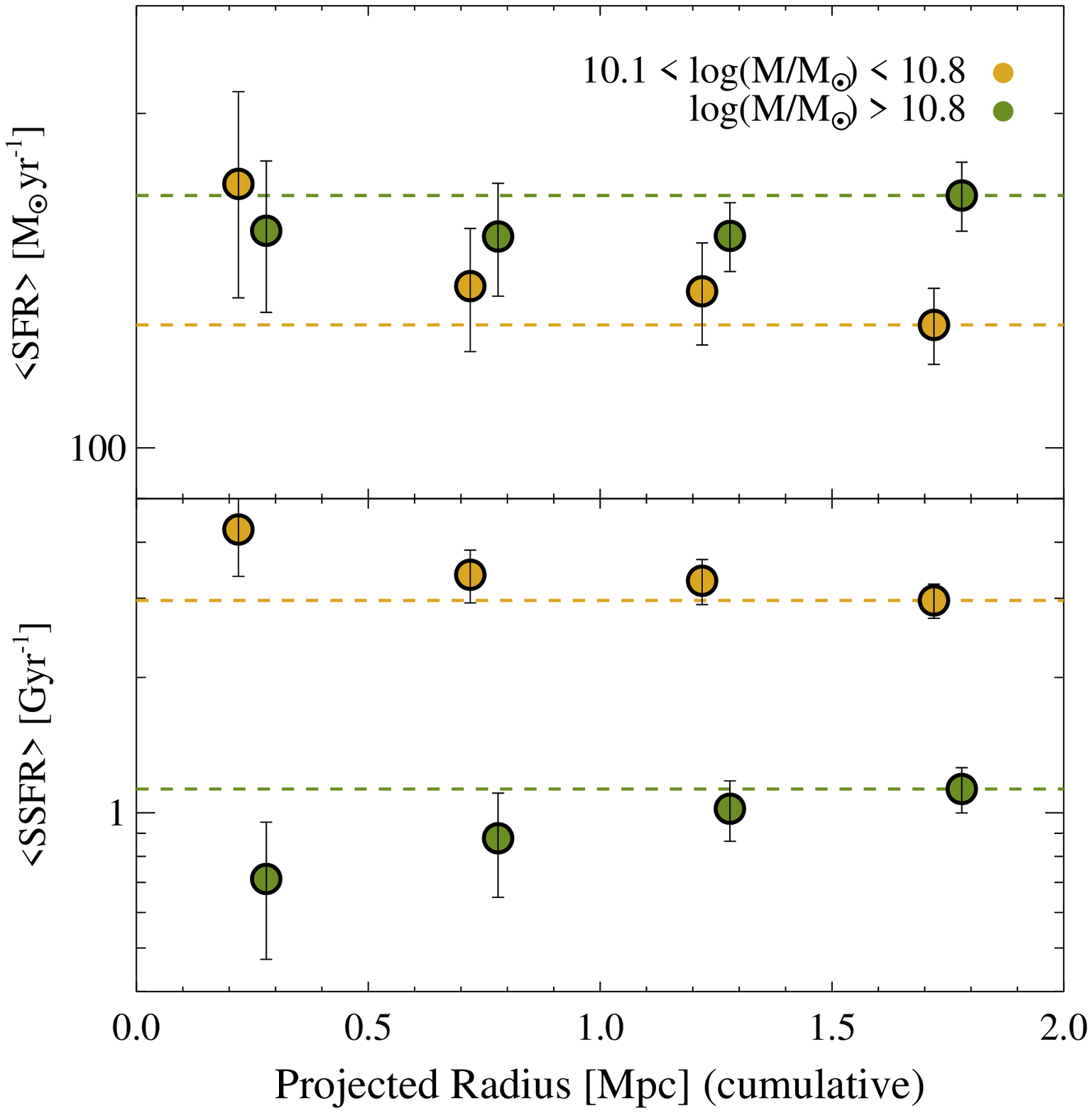}}
\caption{\footnotesize{The average SFR (upper panel) and average SSFR (lower panel) of IR-luminous cluster galaxies as a function of projected radius with galaxies split into two mass bins.  The dashed lines show the representative field values, taken as the outermost radial bin. 
%All quantities are normalized to the field value, taken at a radius of 2 Mpc. 
For higher mass cluster galaxies (log M$_{\star}/\Msun\geq10.8$, green), the $\langle \rm{SFR} \rangle$ is flat with projected radius, while the $\langle \rm{SSFR} \rangle$ is suppressed in the innermost radial bin.  Conversely, the lower mass cluster galaxies (log M$_{\star}/\Msun\leq10.8$, yellow) show a modest increase in both the average SFR and SSFR in the innermost radial bin, indicating enhancement of their SFRs by the cluster environment. All quantities are cumulative with radius.  Errors are determined using bootstrap resampling, which encompasses the spread in the population.}}
\label{fig:sf2}
\end{figure}

In Figure~\ref{fig:sf2}, we break our sample down by stellar mass, in two bins $10.1 < \log (\rm{M_{\star}}/\Msun) <10.8$ and $\log$ (M$_{\star}/\Msun) > 10.8$.    We find that lower mass galaxies show a modest increase in their $\langle\,\rm{SFR}\,\rangle$ and $\langle\,\rm{SSFR}\,\rangle$ into the cluster cores.  Again the errors are determined by bootstrap resampling and encapsulate the spread in the population.  This result supports the idea that any boosting of SF by the cluster environment is primarily apparent in less extreme galaxies, in terms of stellar mass and infrared luminosity, as suggested by our comparison with the \citet{bro13} SFG fraction. The higher mass galaxies [log (M$_{\star}/\Msun) > 10.8$], on the other hand, show a flat $\langle\,\rm{SFR}\,\rangle$ and decreasing $\langle\,\rm{SSFR}\,\rangle$ into the cluster cores.  This decrease is driven by our lower redshift clusters, which we verify by placing this high mass cut on our $1.37<z<1.75$ bin only, which yields a flat trend for $\langle\,\rm{SSFR}\,\rangle$ with projected radius.   We do not have enough cluster galaxies in the cluster cores to split the lower mass sample by redshift and preserve good statistics.

Figures~\ref{fig:sf1} and \ref{fig:sf2} demonstrate that the cluster environment is having multiple effects on the IR-luminous cluster population.  The IR-luminous fraction drops with increasing cosmic time over the redshift range probed, indicating that quenching is becoming effective, and occurring on timescales fast enough to remove IR-luminous galaxies from our sample from $z\sim1.5$ to $z\sim1$.  For galaxies in the cluster cores that remain above our detection limit in both redshift bins, the average SFR is flat with radius, while the average SSFR declines with radius in our lower redshift bin, indicating that the environment is suppressing SFRs relative to stellar mass in excess of what we expect for the evolution of field galaxies on the MS.  This decline is primarily driven by the higher mass galaxies, while the lower mass galaxies show signs of {\it increased} SFRs and SSFRs over what we expect in the field, suggestive of the boosting in the SFG fraction seen in \citet{bro13}.  These results argue for a complex interplay between environment, SFR, and stellar mass, with a general trend from field-like SF to increasingly effective quenching in the IR-luminous cluster population during this epoch. 

\subsubsection{Probing Deeper:  Stacking on the PACS Maps}
\label{sec:stacking}

 In the previous section, we analyzed the IR-luminous cluster population relative to the environment.  Now we look at SF for the full stellar mass-limited cluster population  (log M$_{\star}/\Msun$ $\geq10.1$) using a stacking analysis on the PACS 100$\,\mu$m maps. As each cluster map has a different depth, stacking is performed on each map separately. We combine cutouts centered on the positions of each cluster member and then extract the stacked flux using the same source extraction outlined in Section~\ref{sec:pacsphot}.  Because we are interested in the properties of the full population, we do not remove detected sources from the stack and perform a median stack, which is robust against a small number of bright sources.  The noise within the region we stack is within $\sim15\%$ of the central value, making median stacking more robust than noise-weighted stacking for this analysis. In the case where most sources are near the noise limit, which describes our sample, the results of a median stack are representative of the mean of the population \citep[e.g.][]{whi07}.   Sources are separated by F$_{\rm{gal}}$ during stacking and the final SFRs are obtained through applying the SFG and AGN templates to the appropriate portion of the stacked flux.  

\begin{figure}[!ht]
% Average SFR and SSFR, stacked
\centering
\makebox[\columnwidth]{\includegraphics[angle=270, , trim=10mm 7mm 0 0, clip, scale=0.4]{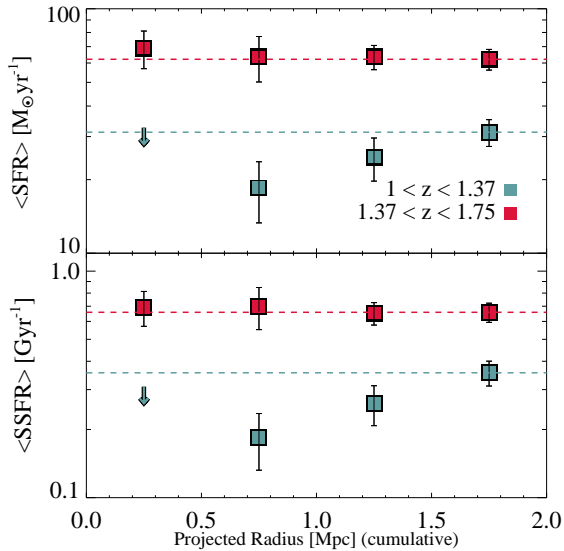}}
\caption{\footnotesize{The average SFR (upper panel) and  SSFR (lower panel) derived from stacking on PACS maps for stellar mass-limited samples of cluster galaxies as a function of projected radius.  As in Figure~\ref{fig:sf1}, the clusters are divided into two redshift bins.  The higher redshift ($1.37<z<1.75$, red) cluster galaxies show flat $\langle \rm{SFRs} \rangle$ and $\langle \rm{SSFRs} \rangle$ into the cluster cores, indicating that the full, stellar mass-limited cluster galaxy sample largely mirrors the SFR and M$_{\star}$ distribution of the field on average.  In the lower redshift clusters (blue), we see a decrease in the $\langle \rm{SFR} \rangle$ and $\langle \rm{SSFR} \rangle$ as quenching and/or mass assembly occurs, in excess of what we expect for field galaxy populations. Errors are the photometric uncertainties of the stack.  Upper limits are $2\sigma$.}}
\label{fig:stack2}
\end{figure}

The combined stacked SFRs and SSFRs of all clusters in two redshift bins can be seen in Figure~\ref{fig:stack2} as a function of projected radius.  In the lower redshift clusters, we see, contrary to the IR-luminous cluster population (Figure~\ref{fig:sf1}), a significant decrease in the average SFR in the cluster cores relative to the field.  Since we are stacking on stellar mass-limited galaxy samples, this is expected due to the increase in the fraction of quiescent and/or quenching galaxies in the cluster environment.  In the higher redshift clusters, however, the stacked average SFR is flat with cluster-centric radius, indicating that the fraction of quiescent or low SFR galaxies does not yet exceed the field value at this epoch.  Similarly, the flat average SSFR indicates that cluster galaxies at $\ga1.4$ are still forming stars for their stellar mass at rates similar to field galaxies at the same epoch.  These trends are fully consistent with the behavior in the PACS-detected cluster galaxy population seen in this work, as well as in \citet{bro13}, and further demonstrates that our cluster sample is undergoing a transition from the epoch of active SF to effective quenching during this era.

We compare these results to the SF properties measured for stellar mass-limited galaxy samples through stacking on SPIRE imaging for the full ISCS cluster sample \citep{alb14}.    The decrease in the average SFR as a function of redshift in this study (Figure~\ref{fig:stack2}) is consistent with the evolution of the average SFR found in that study ($\langle\,\rm{SFR}\,\rangle\sim(1+z)^{5.6}$).  However, \citet{alb14} found that the average SFRs of cluster galaxies at  $\langle\,z\,\rangle=1.2$ are comparable to the stacked $\langle\,\rm{SFR}\,\rangle$ of field galaxies at the same redshifts ($\langle\,\rm{SFR}\,\rangle\sim25\,\Msun$ yr$^{-1}$), while here we find a decrease below the field value at $z\lesssim1.4$.  This apparent discrepancy can be resolved given that the clusters in this work are on the high end (M$_{\rm halo}\gtrsim10^{14}\,\Msun$) of the halo mass distribution of the full ISCS sample.  In addition, the SPIRE stacking analysis found an enhancement in the $\langle\,\rm{SFR}\,\rangle$ of (stellar mass-limited) cluster galaxies over the field at $0.5<r/\rm Mpc<1$ in clusters at $\langle\,z\,\rangle=1.4$, driven by lower mass cluster galaxies.  In this work, we see increased $\langle\,\rm{SFR}\,\rangle$ and $\langle\,\rm{SSFR}\,\rangle$ for the lower mass IR-luminous galaxy population at $r\lesssim0.5\,$Mpc.   We suggest that these differences between the cluster subsample in this work and the analysis of the full ISCS sample demonstrate downsizing effects where more massive clusters, preferentially targeted for additional study in this work, quench SF earlier. 

\subsubsection{Cluster-to-Cluster Variations}
\label{sec:var}

Given that identifying and obtaining multi-wavelength observations of large samples of clusters at $z>1$ is a costly endeavor, multiple studies to date have relied on observations of individual clusters to analyze cluster galaxy populations.  Here we look at variations in star formation from cluster-to-cluster in order to quantify the diversity of high redshift clusters within a uniformly selected sample with similar halo masses.

\begin{figure*}[!ht]
% Stacked and Detected Total
\centering
\includegraphics[angle=270, , trim=0 7mm 0 0, clip, scale=0.55]{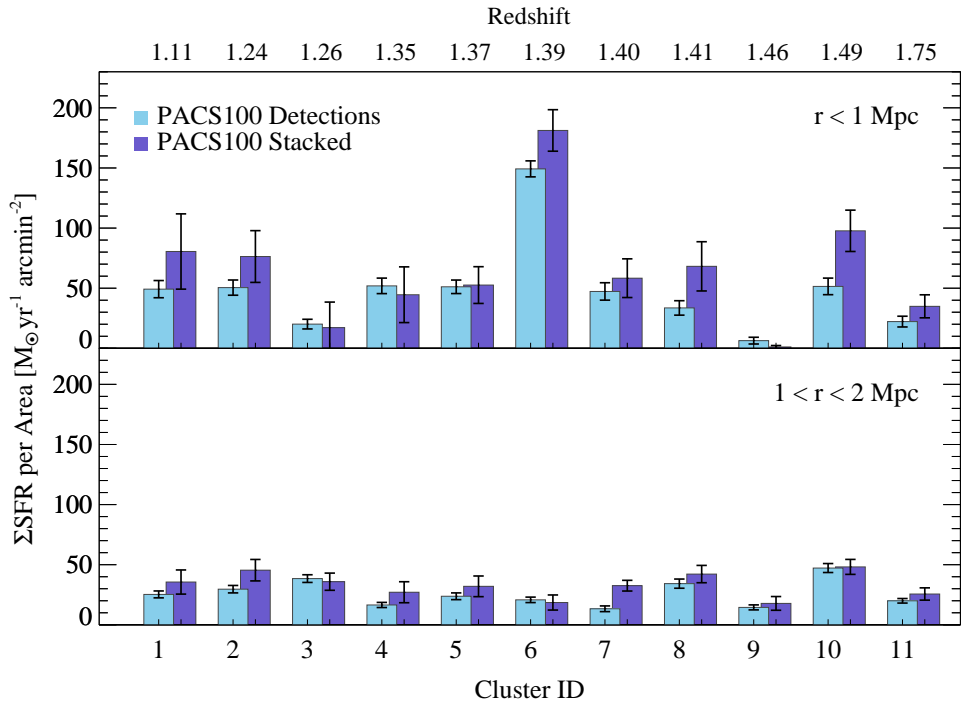}
\caption{\footnotesize{The total SFR per area for each cluster in two radial bins, within the virial radius ($r<1\,\Msun$, top) and in the outskirts ($1<r/\rm{Mpc}<2$, bottom).  Both PACS-detected cluster members (light blue) and stellar mass-limited cluster galaxy samples (dark blue), as determined through stacking, are shown.  Within the virial radius (top), the clusters show a large variation in the total SFR per area, ranging from $\sim\!10-200\,\Msun$ yr$^{-1}$ arcmin$^{-2}$.  In more than half of the clusters, the total SFR from PACS-detected sources is consistent with the total from stacking, indicating that the IR-luminous galaxies are dominating the SFR budget.  This is not true for all the clusters, however, again displaying variations between individual clusters.  Beyond the virial radius, the total SFR per area is much more uniform across our sample, with $\sim20-50\,\Msun$ yr$^{-1}$ arcmin$^{-2}$. } }
\label{fig:sfrperarea}
\end{figure*}

 In the previous section, we looked at the star-forming fraction and average SF properties of IR-luminous and stellar mass-limited cluster galaxy samples (Figures~\ref{fig:sf1}-\ref{fig:stack2}).  The uncertainties in these quantities were determined through bootstrap resampling and thus encompass the intrinsic scatter due to variation in the cluster galaxy populations.  These bootstrapped uncertainties suggest substantial scatter from cluster-to-cluster.  In Figure~\ref{fig:sfrperarea}, we look at the total SFR per area for both PACS-detected cluster members and for stellar mass-limited cluster members, derived by multiplying the stacked $\langle\,\rm{SFR}\,\rangle$ by the total number of stacked sources, for each cluster.  We find that the full range of total SF within the virial radius ($\lesssim1\,$Mpc) spans an order of magnitude ($\Sigma$ SFR per area$\;\sim\,10-200\,\Msun$ yr$^{-1}$ arcmin$^{-2}$), with some clusters showing very little excess SF in the central 1 Mpc while others show up to $\sim7$ times the amount of SFR per area as in the outer 1-2 Mpc.  By contrast, we find less variation at larger radii ($1<r/\rm{Mpc}<2$), with an integrated SFR per area range of $\sim\,20-50\,\Msun$ yr$^{-1}$ arcmin$^{-2}$.  Looking at the outlier-resistant median and interquartile range values (third quartile minus the first quartile, see Table~\ref{tbl:sfstats}), we can see that the interquartile range of the $\Sigma$ SFR per area from detections and stacking in the central 1 Mpc is $1.5-3$ times larger than in the outer radial bin.  To determine if this variation is due to differences in richness from cluster-to-cluster, we repeat this analysis for the total SFR per galaxy for IR-luminous cluster members within the virial radius and at $1<r/\rm{Mpc}<2$.  We find that the interquartile range within the virial radius remains $\sim1.5$ times larger than in the outer radial bin, indicating that we cannot account for the variation in total SF between clusters by richness alone.  The scatter in the total SF is likely due to a combination of the stochasticity of processes such as galaxy mergers and differences in the dynamical states of the clusters.  In addition, the scatter within the virial radius relative to the outskirts suggests that the environmental impact on these galaxies is not dominated by pre-processing among infalling groups, otherwise we might expect similar variation between the two radial bins.  This is consistent with theoretical studies which indicate that pre-processing only becomes important after $z\sim0.5-1$ \citep{mcg09}.

\begin{deluxetable*}{lcccc}
\tabletypesize{\footnotesize}
\tablecolumns{5}
\tablewidth{0pt}
\tablecaption{Statistics - Total Star Formation in Clusters\label{tbl:sfstats}}
\tablehead{
	\colhead{} &
	\colhead{Median} &
	\colhead{First} & 
     \colhead{Third} &
	\colhead{Interquartile} \\
	\colhead{} &
	\colhead{} &
	\colhead{Quartile, Q1} &
	\colhead{Quartile, Q3} &
	\colhead{Range (Q3-Q1)} 
}
\startdata
\multicolumn{5}{c}{} \\
\multicolumn{5}{c}{$\Sigma\,$SFR per area [$\Msun$ yr$^{-1}$ arcmin$^{-2}$]} \\
\multicolumn{5}{c}{} \\
PACS 100$\mu$m Detected Cluster Members ($r<1\,$Mpc) & 49 & 28 & 51 & 23 \\
PACS 100$\mu$m Stacked Cluster Members ($r<1\,$Mpc) & 58 & 40 & 78 & 38 \\
PACS 100$\mu$m Detected Cluster Members ($1<r/\rm{Mpc}<2$) & 24 & 18 & 32 & 14 \\
PACS 100$\mu$m Stacked Cluster Members ($1<r/\rm{Mpc}<2$) & 33 & 26 & 39 & 13 \\
\multicolumn{5}{c}{} \\
\multicolumn{5}{c}{$\Sigma$SFR/M$_{\rm halo}$ [$\Msun$ yr$^{-1}$ per 10$^{14}\,\Msun$]} \\
\multicolumn{5}{c}{} \\
PACS 100$\mu$m Detected Cluster Members ($r<1\,$Mpc) & 207 & 123 & 254 & 131 
\enddata
\end{deluxetable*}

Given the large variation in total SF from cluster-to-cluster, we verify that no single cluster is driving the average trends seen in the previous section by removing each cluster one at a time from our $z>1.37$ bin and recalculating the star forming fraction, average SFR, and SSFR.  We find that the trends in these quantities represent the general cluster sample, though with large variations demonstrated by the bootstrap resampling uncertainties.  

When comparing the total SF within the virial radius from PACS detections versus from stacking, we find a median ratio of 0.8, ranging from 0.6 at the first quartile to a maximum of 1.  We find no strong dependence on redshift for this ratio.   For the majority of our clusters, therefore, the bulk of the SF is occurring in the IR-luminous cluster members with SFR$\,\gtrsim100\,\Msun$ yr$^{-1}$, i.e. our PACS-detected, IR-luminous galaxies are the typical star forming cluster galaxies at this epoch.  We discuss the implications of this further in Section~\ref{sec:disc}.
 
\begin{figure}[!ht]
% Total SFR per halo mass
\centering
\makebox[\columnwidth]{\includegraphics[angle=270, , trim=10mm 7mm 0 0, clip, scale=0.4]{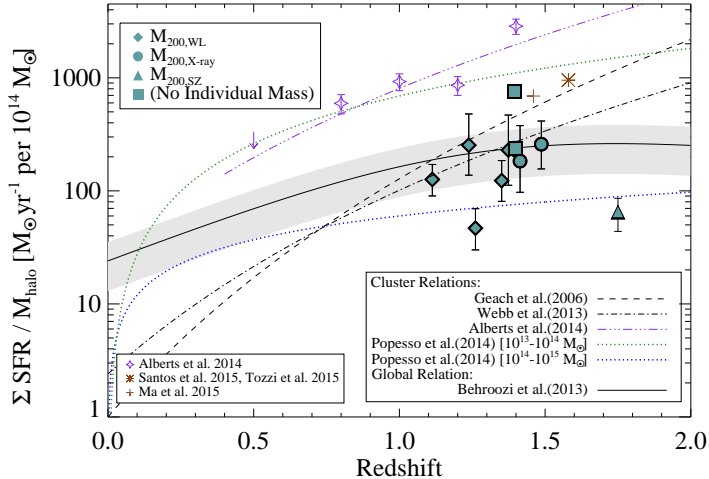}}
\caption{\footnotesize{The halo mass-normalized integrated SFR ($\Sigma$ SFR / M$_{\rm{halo}}$) for IR-luminous cluster members within the virial radius ($\sim\!1\,$Mpc) as a function of redshift.  Halo masses were measured using X-ray (circle), SZ (triangle), or weak lensing measurements (diamond).  Clusters without mass measurements are assigned the median of our X-ray and SZ masses, M$_{\rm{halo}}\sim2.5\e{14}\Msun$ (squares). We compare to three relations measured for clusters at $z\lesssim1$ in the literature from \citet[dashed line;]{gea06}, \citet[dash-dot line;]{web13}, and \citet[blue dotted line;]{pop14} as well as one for massive groups \citet[green dotted line;][]{pop14}.  The purple stars and dash-triple-dot line show $\Sigma$ SFR / M$_{\rm{halo}}$ for the full ISCS cluster sample at $z=0.5-1.5$, assuming a halo mass of $8\e{13}\,\Msun$ \citep{alb14}. In addition, we compare to the global evolution of the mass-normalized SFR for all  galaxies \citep[solid line;][]{beh13}. Also shown are two massive clusters at $z\sim1.5$ with IR observations from previous studies \citep{san15, toz15, ma15}. Our cluster sample generally agrees with the consensus in the literature that the mass-normalized SFR in cluster goes as $(1+z)^{5-7}$ \citep[e.g.][]{gea06, web13}, increasing steeply with redshift and drawing even with the mass normalized total SFR in the field at $z\gtrsim1$.}}
\label{fig:massnorm}
\end{figure}

 In Figure~\ref{fig:massnorm}, we examine the halo mass-normalized integrated SFR ($\Sigma$SFR/M$_{\rm halo}$ where M$_{\rm halo}$ = M$_{200}$) to investigate the relation between SF properties and total mass in clusters.  Halo masses for our clusters are listed in Table~\ref{tbl:clusters}.  Though halo mass estimates are sometimes available from multiple techniques, we adopt mass values from X-ray or SZ observations where available and weak lensing-derived masses otherwise.  Three of our clusters have no independent mass measurement and are assigned M$_{\rm{halo}}=2.5\e{14}\,\Msun$, the median value of our X-ray and SZ mass estimates.  The total SFR is calculated from the sum of the SFRs of the IR-luminous cluster galaxies with log (M$_{\star}/\Msun$) $\geq10.1$ within the cluster virial radius ($r<1\,$Mpc).  We find that, given a relatively small halo mass range of $\sim2-5\e{14}\,\Msun$, the  $\Sigma$SFR per unit halo mass has a full range of about an order of magnitude, with a median of 207 $\Msun$ yr$^{-1}$ per 10$^{14}\,\Msun$ and an interquartile range of 131 $\Msun$ yr$^{-1}$ per 10$^{14}\,\Msun$ (Table~\ref{tbl:sfstats}).  

For comparison, we show the $\Sigma$SFR per unit halo mass derived from IR-luminous galaxies in two massive (M$_{200}\sim3-4\e{14}\,\Msun$) clusters at $z\sim1.5$.  XDCP J0044.0-2033  at $z=1.58$ \citep{san11, toz15, san15} was observed with {\it Herschel}/PACS to have $\Sigma$SFR/M$_{\rm halo} = 954\,\Msun$ yr$^{-1}$ per $10^{14}\,\Msun$ within the virial radius for IR-luminous galaxies with SFR$\;\ga\,165\,\Msun$ yr$^{-1}$.  Observations at 450 and 850$\,\mu$m of XCS J2215.9-1738 at $z=1.46$ \citep{ma15} found $\Sigma$SFR/M$_{\rm halo} = 460^{+210}_{-150}\,\Msun$ yr$^{-1}$ per $10^{14}\,\Msun$ within the virial radius for a SFR limit of $\sim100\,\Msun$ yr$^{-1}$.  These clusters are comparable to our main sample in terms of redshift, mass, and FIR depth; however, they were discovered via their X-ray emission rather than selected in the IR. We find that they have higher star formation for their halo mass than the average of the clusters in this work.  Given the variation we find from cluster-to-cluster within our sample, it is difficult to say whether this difference can be accounted for by different cluster selections.  This comparison emphasizes that analyses of individual clusters are challenging to interpret and that clusters viewed in isolation may bias our understanding of cluster evolution. Measurements of the mass-normalized integrated (infrared) SFR are also available in the literature for Cl 0218.3-0510 at $z=1.62$ \citep{pie12, sma14}; however, we do not show this cluster as its lower mass makes it less comparable to our main cluster sample.

%however, we find that they are much higher than the average of our clusters.  Given the variation we find from cluster-to-cluster, this may indicate that they are not typical of clusters at this redshift and, if viewed in isolation, they may bias our understanding of cluster evolution.  
  
The evolution of $\Sigma$SFR/M$_{\rm halo}$ with redshift has been quantified using cluster surveys in the literature up to $z\sim1$, with some disagreement as to the extrapolation of the behavior of massive haloes at higher redshift \citep[e.g.][]{gea06, koy11, pop12, web13, pop14}.  We compare here to four relations that demonstrate this disagreement, keeping in mind that we have made no effort to correct for differences in cluster selection, galaxy selection, or depth in L$_{\rm IR}$.  \citet{gea06}, examining two clusters at $z\sim0.5$ extrapolated a trend of $\Sigma$SFR/M$_{\rm halo}\sim(1+z)^7$, which they note closely follows the evolution of infrared galaxies in the field up to $z\sim1.5$ \citep[see also][]{cow04}.  A similar result, $\Sigma$SFR/M$_{\rm halo}\sim(1+z)^{5.4\pm1.5}$, derived for IR-luminous cluster galaxies up to $z\sim1$ for $\sim3\e{14}\,\Msun$ haloes was found by \citet{web13}. As seen in Figure~\ref{fig:massnorm}, these trends indicate that star formation in massive clusters is at or suppressed below the field at $z<1$.  However, at higher redshifts, cluster SF may surpass SF in lower mass haloes, indicating a true reversal in the SFR-density relation \citep[see also e.g.][]{tra10, hil10, bro13, alb14, san15}.  By contrast, \citet{pop14}, using X-ray selected clusters up to $z\sim1$ ($10^{14-15}\,\Msun$, blue dotted line) and massive groups ($10^{13-14}\,\Msun$, green dotted line) up to $z\sim1.5$, predict a flattening in $\Sigma$SFR/M$_{\rm halo}$, with star formation in the most massive haloes always suppressed below the global, field level.  This can be seen in Figure~\ref{fig:massnorm}, where the field value is represented by the solid line and gray shaded region which show the evolution of the $\Sigma$SFR/M$_{\rm halo}$ of all galaxies up to $z\sim2$, quantified as the observed SFR density presented in \citet{beh13} divided by the mean comoving matter density of the Universe.  We note that this global relation is formally a lower limit, as we have made the assumption that all matter is locked in haloes that host galaxies.  In actuality, the fraction of matter locked in occupied haloes depends on the local density field and so our mass-normalized integrated SFR for the field may be underestimated by up to a factor of 2-3 \citep{fal10}.  Since this factor is fairly uncertain, we do not correct for it.

In comparison with these relations, cluster and field, we see that the typical $\Sigma$SFR/M$_{\rm halo}$ for our PACS clusters is largely consistent with that in the field, as represented by the global relation, over the redshift range probed here. Our clusters are also consistent with the redshift evolution of the $\Sigma$SFR/M$_{\rm halo}-z$ relation found by \citet{gea06} and \citet{web13}; however they are inconsistent with SF being suppressed below the global field SFR in $\gtrsim10^{14}\,\Msun$ haloes at all redshifts as suggested in some studies \citep[i.e.][]{pop14}.  

Lastly, in Figure~\ref{fig:massnorm}, we show the evolution of the mass-normalized integrated SFR for the all ISCS clusters from $z=0.5-1.5$.  Using the average SFRs derived from stacking on {\it Herschel}/SPIRE imaging and assuming a typical halo mass of $8\e{13}\,\Msun$ that is constant with redshift \citep[see][for details]{alb14}, we find a best-fit function of $\Sigma$SFR/M$_{\rm halo}\sim(1+z)^{4.9\pm1.1}$  for mass-limited cluster galaxy samples from the full ISCS.  The redshift evolution of the ISCS clusters, as represented by the shape and slope of this function, is therefore in good agreement with the redshift evolution up to $z\sim1$ as measured in \citet{web13}, despite differences in cluster and galaxy selection.  On the other hand, we see a significant disparity in the amplitude of $\Sigma$SFR/M$_{\rm halo}$ for the full ISCS sample, which falls well above that measured for the \citet{web13} clusters and our PACS-selected cluster sample.  We reiterate that we have made a simplifying assumption by adopting the median mass of the ISCS as measured statistically in \citet{alb14} in this analysis, which will affect the normalization.  Keeping this assumption in mind, we attribute this difference in normalization to two main factors.  The first is cluster selection: the ISCS clusters are selected in the rest-frame NIR, which probes the {\it in situ} stellar mass content, whereas the Red Sequence Cluster \citep[RCS;][]{gla05} sample used in \citet{web13} were selected using an observable that depends on a cluster’s star formation history.  Specifically, at fixed mass, the RCS is more sensitive to clusters with higher passive galaxy fraction, and, correspondingly, more prominent and tighter red sequences.  This may, in part, explain why the overall level of SF is lower in the RCS sample.  The difference between the full ISCS sample and the ISCS/IDCS clusters observed with PACS, however, suggests a second factor, the halo mass, plays at least a partial role.  The $\Sigma$SFR/M$_{\rm halo}$ - M$_{\rm halo}$ relation in clusters is still uncertain.  Current estimates suggest a strong link between star formation in clusters and halo mass, with \citet{web13} finding $\Sigma$SFR/M$_{\rm halo}\sim\;$M$_{\rm halo}^{-1.5\pm0.4}$ for clusters up to $z\sim1$ using richness as a proxy for halo mass \citep[see also][]{pop14}.  The difference in amplitude between the full ISCS sample and the higher mass PACS-selected subsample is broadly consistent with this relation (though see above for a discussion on the cluster-to-cluster variation in SF activity relative to cluster richness for the PACS-selected sample).  A large, uniformly-selected cluster survey with well characterized halo masses and star formation activity is needed to further constrain the $\Sigma$SFR/M$_{\rm halo}$ - M$_{\rm halo}$ relation.

\subsection{The Epoch of AGN Activity in Cluster Cores at $1<z<2$}
\label{sec:agn}

Studies of X-ray, MIR, and radio-selected AGN in the ISCS cluster sample have established that the AGN fraction increases dramatically within cluster environments to high redshift, climbing to field-like AGN fractions at $z\sim1.25$ \citep{gal09, mar09, mar13}, a two order of magnitude increase over local clusters.  In this section, we examine the evolution of the AGN fraction in clusters selected through SED fitting as a function of redshift and radius.  Using SED fitting allows us to identify lower luminosity AGN than can be selected through shallow-to-moderate depth X-ray/radio observations or MIR color diagnostics (see Figure~\ref{fig:fgal}). To avoid the bias against luminous AGN introduced by requiring photometric redshifts for cluster membership, for this analysis we opt to do a line-of-sight study in order to isolate cluster trends. We therefore analyze all galaxies along the line-of-sight to our clusters regardless of the photometric redshift of the galaxy. To increase our statistical power and take advantage of all available data, we expand this analysis to include $\sim250$ ISCS plus three IDCS clusters between $0.5<z<2$. 

We again divide the galaxies in the photometric redshift catalog into four categories: F$_{\rm{gal}}<0.3$ (``AGN-dominated"), $0.3<\;$F$_{\rm{gal}}<0.5$ (``AGN-composite"), $0.5<\;$F$_{\rm{gal}}<0.7$ (``host-composite"), and F$_{\rm{gal}}>0.7$ (``host-dominated").  Note that the F$_{\rm{gal}}$ parameter is not sensitive to star formation activity, so these categories should not be interpreted as SFGs versus AGN, but rather by a decreasing degree of AGN influence on the UV-MIR SED of all galaxy types, including non-star forming ellipticals.    

\begin{figure}[!ht]
\centering
\makebox[\columnwidth]{\includegraphics[angle=270, width=0.9\linewidth, trim=0 5mm 0 0, clip]{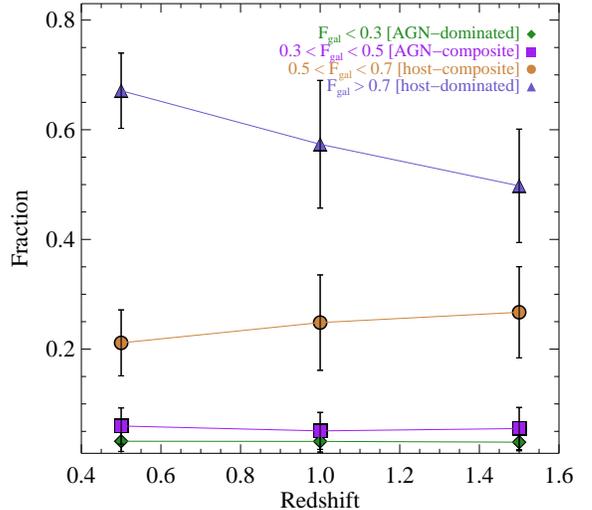}}
\caption{\footnotesize{The fraction of galaxy subtypes along the line-of-site ($r<0.5\,$Mpc) to 248 ISCS/IDCS clusters from $z=0.5-2$.  Galaxies are separated into four subtypes by the F$_{\rm{gal}}$ parameter: AGN-dominated (F$_{\rm{gal}}<0.3$, green diamonds), AGN-composite ($0.3<\rm{F}_{\rm{gal}}<0.5$, purple squares), host-composite ($0.5<\rm{F}_{\rm{gal}}<0.7$, yellow circles), and host-dominated (F$_{\rm{gal}}>0.7$, blue triangles). Though host-dominated sources make up the bulk of sources at all redshifts, their fraction decreases with increasing redshift in the cluster cores. }  }
\label{fig:agn1}
\end{figure}

Figure~\ref{fig:agn1} shows the weighted average of the fraction of each category along the line-of-sight to the ISCS/IDCS cluster cores ($r<0.5\,$Mpc) as a function of redshift.  Host-dominated galaxies dominate the numbers; however, we find a marked decrease in this subtype with increasing redshift, from 65$\%$ to 48$\%$ from $z = 0.5$ to $z=1.5$.  The bulk of this difference is countered by an increase in host-composite galaxies, with smaller gains in the AGN-composites and AGN-dominated galaxies. 

\begin{figure*}
\centering
\includegraphics[angle=270, width=0.7\linewidth]{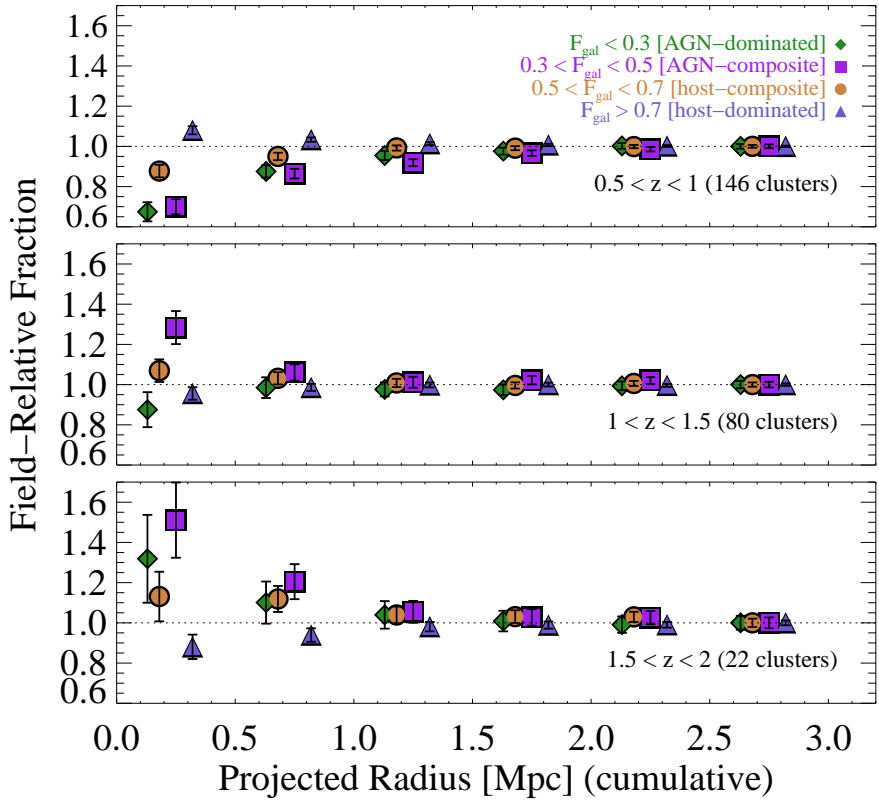}
\caption{\footnotesize{The fraction of each galaxy subtype as a function of projected cluster-centric radius, normalized to the field value at 3 Mpc. The ISCS/IDCS clusters are separated into three redshift bins: $0.5<z<1$ (upper panel), $1<z<1.5$ (middle panel), $1.5<z<2$ (lower panel).  For the lowest redshift clusters, host-dominated sources are present in the cluster cores in slight excess of the field, while composite and AGN-dominated galaxy fractions are below the field level, indicating that AGN activity is no longer being fueled in these clusters.  These trends reverse with increasing redshift, and for clusters at $1.5<z<2$ we see significantly enhanced fractions of AGN-dominated and composite sources in relation to the field, indicating that the cluster environment is triggering the growth of AGN in cluster galaxies, likely through galaxy interactions such as mergers.}}
\label{fig:agn2}
\end{figure*}

Next we examine the field-relative fraction of each galaxy subtype as a function of cluster-centric radius in order to isolate how much of the evolving fraction is due to the cluster environment.  We bin our clusters into three redshift bins: $0.5<z<1$ (146 clusters), $1<z<1.5$ (80 clusters), and $1.5<z<2$ (22 clusters).  We then quantify the fraction of each subtype along the line of sight in radial bins, out to projected radii of 3 Mpc, which is taken to be the field value (Figure~\ref{fig:agn2}).  We adopt 3 Mpc rather than the 2 Mpc used in the SF analysis as we are not limited by the PACS footprint and therefore can go to larger radii to look for variations in the cluster outskirts.  We verify that normalizing at 3 Mpc over 2 Mpc does not chance our results.  Nor does using cumulative or differential annuli, which indicates that there is no significant variation beyond 1 Mpc due to environment.  In the lowest redshift bin, host-dominated galaxies are overrepresented in clusters at $\sim110\%$ of the field value, with host/AGN-composites and AGN-dominated galaxies underrepresented by $\sim10-30\%$.  By $z=1$, host-dominated galaxies have dropped below the field level, with host-composites slightly above and AGN-composites rising to $\sim130\%$ of the field level.  The fraction of AGN-dominated galaxies has also risen, though it is still below the field value.  For our highest redshift clusters, however, AGN-dominated, AGN-composites, and host-composites are all above the field, with AGN-composites at $150\%$ of the field level.  

These results show a substantial rise in the fraction of AGN and AGN-composites in the cluster cores, consistent with previous studies \citep{gal09, mar09, mar13}.  In addition, we demonstrate a rise in AGN-composite galaxies, representing relatively weak AGN and/or strong host galaxies, and a decline in host-dominated sources with $<30\%$ contribution to their UV-MIR SED from AGN emission.  The implications of this and how it relates to the observed increase in star formation with redshift is discussed in Section~\ref{sec:disc}.

\section{Discussion}
\label{sec:disc}

\subsection{Variations Between Individual High Redshift Clusters}

Due to observational challenges and the increasing scarcity of massive clusters at high redshifts, detailed studies of cluster populations during this epoch are typically performed on individual clusters.  It is therefore important to understand the range in variation from cluster-to-cluster in the high redshift ($z>1$) regime.    In this work, we have analyzed a statistical sample of uniformly selected clusters with similar halo masses over a relatively small redshift range using self-consistent techniques for identifying cluster membership and measuring cluster galaxy properties.  We find a significant scatter in the total dust-obscured star formation activity across our cluster sample.  This is best seen in Figure~\ref{fig:sfrperarea}, where the total SFR per area covers a large range, with greater variation in the cluster cores than in the outskirts ($1<r/\rm{Mpc}<2$).  Strong variations in the mass-normalized integrated SFRs of massive clusters at $z\sim0.5$ were first noted in \citet{gea06}; we similarly find a wide range in $\Sigma$SFR/M$_{\rm halo}$ for our cluster sample.  This scatter in the SF activity at roughly fixed halo mass introduces the possibility of significant bias in single cluster studies.   Differences in the SF activity of cluster populations is likely due to a variety of factors, including differences in assembly history and dynamical state. As noted in \citet{muz12}, even relatively evolved, rich clusters at $z\sim1$ show a wide range of morphologies.  The X-ray observations available for a few clusters (ID8, ID10, ID11) in our sample indicate that these clusters are unrelaxed, with associated filamentary structures, and, in one case, may have undergone a recent interaction or merging event \citep{bro11, bro15}.  As clusters at higher redshift are more likely to have recently experienced evolution in their dynamical state, it becomes more and more important to work with statistical cluster samples which can be used to average over these variations.

Of particular note in our sample are the clusters ID6 ($z=1.396$) and ID10 ($z=1.487$).   As described in Section~\ref{sec:membership}, these two clusters have a small angular separation ($\sim4^{\prime}$ or $\sim2\,$Mpc at $z\sim1.4$), and given their overlap both spatially and in photometric redshift space, we have assured no double-counting of galaxies by assigning membership based on the maximum integrated PDF of the photometric redshifts.   As a system and individually, these two clusters stand out among our sample, with  $\Sigma$SFR per area 2-9 times larger within the virial radius than in the surrounding  outskirts.  ID6 and ID10 have the highest fractions of PACS-detected, IR-luminous galaxies within 250 kpc in our sample, $\sim60\%$ and 50$\%$, respectively, relative to a median fraction of $\sim20\%$ for the full cluster sample.  Interestingly, ID10 also has nearly half of its SF coming from galaxies below our PACS detection limit, as determined through stacking.  Evidence suggests that these lower SFR galaxies may be undergoing enhancement by the cluster environment (see Section~\ref{sec:sfradius} and Section~\ref{sec:var}).  Though their separation in redshift space (nearly 200 Mpc, comoving line of sight) makes it unlikely these clusters are currently merging, we speculate their substantial star formation may be related to their connection within large scale structure. If so, then these clusters represent an important example of the effects of the dynamical state on cluster galaxies at high redshift, complimentary to lower redshift studies of more extreme merging cluster systems \citep[e.g.][]{clo04, chu10, men12}.  Spectroscopic follow-up of this system to more accurately map cluster membership and localize SF to the clusters and any filamentary structure in between would provide important constraints on the hierarchical growth of clusters at high redshift.  

We also highlight ID11, the most massive cluster known at $z>1.5$.  ID11 \citep{sta12} is detected in both SZ \citep{bro12} and X-ray \citep{bro15}, with a consistent ICM-based mass of M$_{200}\approx4\e{14}\,\Msun$ \citep[see also][for a weak lensing analysis]{mo16}.  As demonstrated in \citet{bro12}, given its mass and redshift ($z=1.75$), ID11 is a true predecessor to Coma-like clusters, the most massive virialized structures in the local Universe. It is therefore unique within our cluster sample.  Recently, deep X-ray measurements have also determined that ID11 is possibly a cool-core cluster that has experienced a recent interaction or merger \citep{bro15}.  

As discussed in \citet{bro13}, the transition epoch from active star formation to efficient quenching is predicted to occur earlier for more massive systems.  Evidence for such cluster downsizing \citep[e.g.][]{nei06} was presented in \citet{wyl14}, which looked at the MIR luminosity function of cluster candidates around radio-loud AGN at $z\sim1-3$.  In contrast with the \citet{man10} results discussed earlier, they found that the luminosity function of cluster candidate galaxies were consistent with passive evolution models and high formation redshifts ($z_f\sim3$).  Given that radio-loud AGN are expected to reside in extremely massive haloes, these two studies are consistent within the framework of cluster downsizing.  Further corroborating this finding, \citet{wel16} recently found significant star formation in the environments of $\sim10^{13}\,\Msun$ haloes at $z\sim1$, in sharp contrast with the quenched populations of more massive clusters at the same redshift  \citep[e.g.][]{pat09, bro13, alb14}. We therefore might expect ID11 to have undergone an earlier transition epoch and to be relatively evolved, given its redshift and mass.  Indeed, we find that ID11 has little SF activity in its core ($\sim15\%$ detection fraction with PACS at $<250\,$kpc) and a low mass-normalized SFR (Figure~\ref{fig:massnorm}) compared to the rest of our sample and the global field relation, indicating it is already an evolved system.  We note, however, that there is a similar mass cluster known at $z=1.58$ that shows substantial ongoing SF activity \citep{toz15, san15}, again stressing that the scatter due to cluster-to-cluster variations limits what we can conclude based on individual clusters.

\subsection{The Co-Evolution of Star Formation and AGN in Clusters}

To first approximation, the co-evolution of star formation and black hole (BH) growth (i.e.\,AGN activity) seems unsurprising, given that both processes are driven by the availability of the same cold gas supply.  The disparate size scales, however, with SF occurring throughout in the disk and AGN growth at sub-kpc scales, make establishing a link between these two processes challenging \citep[see][for reviews]{ale12,bra15}.  Simulations find that the physical processes that feed BH growth on small spatial scales are unlikely to be smooth or continuous, leading them to vary dramatically on short timescales \citep{hop10, hic14}. This variation may hide a strong underlying correlation with longer-lived SF activity \citep[e.g.][]{gab13, hic14, nei14, tha14, del15, xu15}.   The relation between average BH growth and global SFRs, combined with the parallel redshift evolution of SF and AGN activity \citep[see][for a review]{mad14} and possible observations of an AGN Main Sequence \citep{mul12b}, suggest an important link, the nature of which is still heavily debated.  AGN can be triggered by either internal secular evolution processes$-$disk instabilities, bars, etc$-$or through galaxy interactions such as harassment and minor and major mergers \citep[see][for a review]{ale12,bra15}, which may be an important mechanism for BH growth in overdense environments \citep[e.g.][]{bro13, ehl15}. 

In this work, we looked at the field-relative fraction of AGN-dominated and AGN-composite cluster galaxies as a function of redshift in a sample of $\sim250$ ISCS/IDCS clusters from $0.5<z<2$ (Figure~\ref{fig:agn2}).  We found that the fraction of AGN, as selected through SED fitting, increases steeply in cluster cores with increasing redshift, consistent with previous studies of luminous AGN  \citep[e.g.][]{gal09, mar13}.  The fraction of AGN-dominated and AGN-composite cluster galaxies is below the field level at $z<1$, consistent with previous studies \citep[e.g.][]{gal09, ehl13, ehl14}.  At $z=1-1.5$, AGN-dominated cluster galaxies increase to just below field level, while AGN-composites are found in excess of the field.   This rise in the AGN-composite fraction indicates that the cluster environment stimulates moderate BH growth in addition to the luminous AGN systems previously indicated by X-ray, radio, and MIR selections.  By $z>1.5$, both AGN-dominated and AGN-composite systems are found in excess of the field level, implying that the environment is triggering AGN over a range of luminosities through increased galaxy interactions in $\sim10^{14}\,\Msun$ haloes.  The rapid decline in excess AGN-dominated galaxies by $z\sim1-1.5$ suggests a transition toward fewer galaxy interactions as the cluster virializes and cluster velocity dispersions increase.  This transition epoch is remarkably similar to the behavior we see for the evolution of SF activity in cluster galaxies, wherein SF quenching becomes effective around the same epoch. AGN could provide a mechanism for quenching SF by heating and/or expelling cold gas.  In addition to permitting a rapid evolution of cluster galaxies onto the red sequence \citep[see][]{bro13}, this feedback would also lead to the suppression of AGN activity in clusters at lower redshifts.

\subsection{Star Formation and the Cluster Environment}

Star formation in cluster galaxies can be influenced by multiple mechanisms specific to overdense environments, the efficiencies of which likely vary as clusters grow in mass and/or undergo virialization.  Some of the commonly invoked mechanisms include strangulation \citep{lar80} $-$ the removal of loosely-bound hot halo gas by the intracluster medium (ICM) on long timescales $-$ ram pressure stripping \citep[RPS;][]{gun72} $-$ the removal of the interstellar medium through interactions with the ICM on moderate timescales $-$ and galaxy interactions such as harassment \citep{moo96} and/or mergers.  For an overview of these processes, see \citet{bos06}.  

Recently, studies have revealed that galaxies in the field at high redshift ($z=1-3$) have short gas depletion timescales  \citep[$\sim0.7\,$Gyr; ][]{dad10, tac10,mag13}, stressing the connection between new gas accretion and prolonged SF.   This has important implications for overdense environments where the hot ICM in clusters will suppress new gas accretion from cold flows  \citep[e.g.][]{ker05} and where processes such as strangulation and RPS may limit available gas supplies on distinct timescales.  

In Section~\ref{sec:var}, we found that IR-luminous cluster members are the typical SFGs in our clusters, accounting for the bulk of cluster SF.  We consider this result in conjunction with: 1) the flat $\langle\,\rm{SFR}\,\rangle$ of IR-luminous cluster galaxies as a function of cluster-centric radius (Figure~\ref{fig:sf1}) for both of our redshift bins, and 2) our stacking analysis (Figure~\ref{fig:stack2}), which shows a marked decrease in the $\langle\,\rm{SFR}\,\rangle$ at small projected radii at $z<1.37$ for stellar mass-limited cluster samples.  Taken together, these lines of evidence indicate that these clusters are  undergoing a rapid build-up of the red sequence, with star forming galaxies experiencing significant quenching over short timescales (the period spanning the median redshifts of our bins, z=1.46 to z=1.26, is $\sim0.6$ Gyr), rather than a gradual decrease in the SFRs of cluster SFGs over long timescales.  Further constraining the timescale to fully quench cluster galaxies will require measurements of the star formation histories in individual cluster galaxies, both star forming and quenched, over a long redshift baseline.

Strangulation typically occurs over several Gyr, too long to cause the transition seen in the star forming fraction of cluster galaxies at $z\sim1.4$, indicating that it is not dominating quenching in high redshift clusters. RPS is a more rapid process, expected to act on a timescale less than or similar to the crossing time \citep[$\sim1-2\,$Gyr for our cluster sample, assuming velocity dispersions of 700 km s$^{-1}$;][]{bro11}.   However, though examples of efficient RPS have been observed in massive clusters locally \citep[e.g.][]{ebe14}, RPS fails to explain the enhanced SF in lower mass galaxies and excess AGN fraction in the cluster cores that we find in this work as well as the rapid reddening of cluster galaxies colors by $z\sim1$ \citep{eis08}.  As described in \citet{alb14}, however, we found that quenching continues at later epochs ($z<1$) at rates faster than in the field, with an average timescale consistent with strangulation and RPS, and we posit that these processes may also be related to the decline in AGN activity below what is observed in the field \citep[e.g.][this work]{gal09}.  A vital next step in evaluating the evolution of cluster populations from low to high redshift is directly observing the ISM, particularly the molecular gas supply, in cluster galaxies in relation to SF and AGN activity.  Measurements of the mass of the ISM in cluster galaxies at $z=1.75$ will be presented in future work (Alberts et al., in preparation). 

There is growing evidence that galaxy interactions, specifically mergers, play a dominant role in clusters at high redshift, where the conditions for galaxy interactions are favorable due to high galaxy space densities and low relative velocities \citep[$\sim700\,$km s$^{-1}$ in the ISCS clusters; ][]{bro11}.  For example, rapid mass growth in cluster galaxies at $z\gtrsim1.3$ \citep[e.g.][]{man10, man12} and a dearth of massive red sequence galaxies at high redshift \citep[e.g.][]{fas11,rud12,man12} provide statistical evidence of merger activity.  An enhanced merger fraction has been  observed in a cluster at $z=1.62$ \citep{lot11}, and cluster ETGs at $z>1$ have been found to have stochastic star formation histories \citep{sny12} and residual star formation \citep{wag15,mei15}, indicating that they recently and rapidly quenched.  A significant fraction of cluster SFGs and ETGs have been observed to have disturbed morphologies in $z>1.5$ clusters \citep{san15, mei15}. Minor and/or dry mergers may additionally explain the larger size of quiescent galaxies in clusters at high redshift relative to the field \citep{pap12, bas13, del14, str15}.  

In this work, we have demonstrated an excess in the AGN fraction, which we attribute to increased galaxy interactions, such as mergers.  In addition, we find that the average SSFR of IR-luminous cluster galaxies at $z\lesssim1.4$ is suppressed in the cluster cores, while their average SFRs remain constant.  This suggests mass build-up which occurs without significantly altering SFRs, either through enhancement or quenching, which may indicate dry merger activity.  Together with previous studies, these lines of evidence indicate an enhanced merger rate.  For further discussion of the role of mergers and AGN in quenching SF in high redshift clusters, see \citet{bro13}.

\section{Conclusions}

In this work, we have examined the star formation properties of galaxies in 11 spectroscopically confirmed massive ($\sim10^{14}\,\Msun$) clusters at $z=1-1.75$ from the ISCS/IDCS.  Using new deep {\it Herschel}/PACS imaging at 100 and 160$\,\mu$m, we have characterized the obscured star formation in IR-luminous cluster galaxies, including those hosting AGN (identified through optical-to-MIR SED fitting).  We present the first optical-FIR SEDs of high redshift cluster galaxies, quantify robust SFRs and SSFRs and characterize these quantities in terms of cluster-centric radius, redshift, and halo mass.  Stacking is used to compare the IR-luminous cluster population to star formation in mass-limited cluster galaxy samples.  We highlight both trends within our full cluster sample and variations between individual clusters.  Finally, the fraction of AGN in clusters from $z=0.5$ to $z=2$ is evaluated.  Our results are as follows:

\begin{itemize}
\item[i.] The near- to far-infrared SEDs of cluster galaxies at high redshift can be well described, on average, by empirically derived templates for SFGs and AGN in the field from \citet[][]{kir12, kir15} with similar infrared luminosities and redshifts.  This result indicates that field galaxy templates can be used to derive robust cluster SFRs from infrared observations.  Further analysis of the far-infrared SED and dust properties of cluster galaxies at $z=1.75$ using submillimeter observations will be presented in Alberts et al. (in preparation).

\item[ii.] The star forming fraction and average SFRs and SSFRs of IR-luminous cluster galaxies as a function of cluster-centric radius indicate a transition from field-like SF activity at $z\ga1.4$ to significant quenching at lower redshift, consistent with previous studies \citep{bro13, alb14}.   We find both a significant reduction in the star forming fraction and the $\langle\,\rm{SSFRs}\,\rangle$ from $z\sim1.5$ to $z\sim1$, indicating that some IR-luminous galaxies are being quenched below our detection limit while others are being prevented from forming stars at the same rate as galaxies found in the field {\it for their mass}.  When split by stellar mass, our cluster sample shows evidence for enhanced SF activity in lower stellar mass ($10.1<\log(\rm M/\Msun)<10.8$) cluster galaxies, as has been seen in previous studies \citep{bro13, alb14}. Stacking on mass-limited cluster samples demonstrates that IR-luminous galaxies dominate the SFR budget in high redshift clusters.

\item[iii.] Galaxy clusters in our uniformly selected sample show a significant variation in their star formation properties from cluster-to-cluster.  The total SFR per area within the virial radius ($r<1\,$Mpc) ranges by an order of magnitude between clusters of similar halo mass and at similar redshifts.  This is in sharp contrast to the modest (factor of $\sim2$) range in the cluster outskirts ($1<r/\rm{Mpc}<2$). We examine the mass-normalized total SFR as a function of redshift, finding that our clusters are largely consistent with the level of SF activity in the field over the redshift range probed and are consistent with the $\sim(1+z)^{5-7}$ redshift evolution determined for $z<1$ clusters \cite[e.g.][]{gea06, web13}.   Scatter in the total SF activity from cluster-to-cluster highlights the need for large, uniformly selected cluster samples and cautions against over-interpretation in studies of individual clusters.

\item[iv.] Using $\sim250$ ISCS/IDCS clusters from $0.5<z<2$, we examine the fraction of AGN as a function of cluster-centric radius and redshift. We quantify AGN content by the contribution of AGN emission to the optical-to-MIR SED, a long wavelength baseline which allows us to select a broad range of AGN types \citep[e.g.][]{hic09, men13, chu14}.  We find that AGN-dominated and AGN-composite galaxies are found in excess of the fraction in the field in high redshift ($z\gtrsim1$) clusters, indicating that the cluster environment is fueling AGN through galaxy interactions.  The decline in the AGN fraction parallels (and possibly precipitates) the decline in the IR-luminous cluster population, suggesting a co-evolution between black hole growth and SF activity in overdense environments.
\end{itemize}

\acknowledgments

The authors thank their colleagues in the IRAC Shallow/Distant Cluster Survey, IRAC Shallow Survey, NDWFS, SDWFS and MAGES teams, in addition to the HerMES collaboration for making their data publicly available.  This work is based on observations made with {\it Herschel}, a European Space Agency Cornerstone Mission with significant participation by NASA. Support for this work was provided by NASA through an award issued by JPL/Caltech. The authors extend a special thanks to Bruno Altieri and Hanae Inami for their assistance in {\it Herschel} data reduction and source extraction and to Ranga-Ram Chary for discussions on the source extraction and analysis.  HIPE is a joint development by the {\it Herschel} Science Ground Segment Consortium, consisting of ESA, the NASA {\it Herschel} Science Center, and the HIFI, PACS and SPIRE consortia.  This work is additionally based on observations made with {\it Spitzer}, which is operated by the Jet Propulsion Laboratory, California Institute of Technology under contract with NASA.  Finally, the authors thank the anonymous referee for their constructive and helpful comments which have improved this work.

\appendix
 \setcounter{figure}{0} \renewcommand{\thefigure}{A.\arabic{figure}}
\setcounter{table}{0} \renewcommand{\thetable}{A.\arabic{table}}

\section{PACS Maps: Description and Monte Carlo Simulations}
\label{appendix:a}

Imaging at 100 and 160$\,\mu$m is available for 11 spectroscopically confirmed clusters from Open Time 2 observing program OT2\_apope\_3.  Nine of the clusters are observed in individual maps and a tenth map contains two clusters (ID6 and ID10) due to their small angular separation.  Integration times range from 270 to 4050 s, providing uniform sensitivity to IR-luminous galaxies for each cluster from $z=1-1.75$.   Each map is observed with at least two AORs with two different scan directions, offset by 90 degrees, in order to remove stripping effects from the 1/f noise.  The observation IDs for each map can be seen in Table~\ref{tbl:a}.  Each map covers an area of 7$^{\prime}$x7$^{\prime}$ with uniform sensitivity in the central 5$^{\prime}$x5$^{\prime}$.  

Data reduction and source extraction are performed as described in Section~\ref{sec:pacs}.  PSF fitting at the location of priors is done using the empirical PSF derived from observations of the Vesta asteroid.  To remove excess noise in the PSF wings, we truncate the 100$\,\mu$m (160$\,\mu$m) Vesta PSF to a size of 6 (5) pixels and apply an aperture correction of 0.660 (0.705) to extracted sources.  The rms sensitivities range from $\sim$0.5-2 mJy, based on extracting the flux from 10,000 randomly placed apertures on the residual maps.  

\begin{figure}[!ht]
\makebox[\columnwidth]{\includegraphics[angle=270, width=0.8\linewidth]{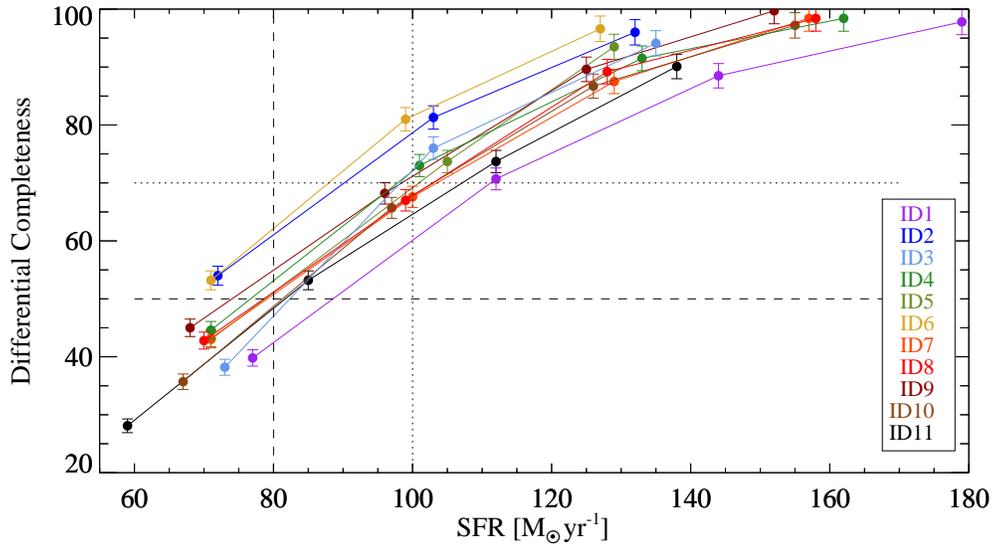}}
\caption{\footnotesize{The differential completeness in the central 5$^{\prime}$x5$^{\prime}$ as a function of SFR, where the SFR corresponds to different flux densities depending on the redshift of the cluster, assuming an empirical SFG template \citep{kir12}.  The dashed (dotted) lines indicate the 50$\%$ (70$\%$) completeness level. }}
\label{fig:completeness}
\end{figure}

Monte Carlo simulations are performed on the 100$\,\mu$m maps to assess the completeness of each map and the photometric accuracy and noise properties of extracted sources.  Simulated sources are inserted into the signal maps at discrete flux levels using the 100$\,\mu$m Vesta PSF.  In order to preserve the original map statistics, 20 simulated sources at a given flux density are inserted at a time and the process is repeated for a total of 5,000 simulated sources per map per flux bin.  Flux bins are chosen based on the depth of each map such that we test the completeness and photometric accuracy down to uniform limits in SFR at the redshift of the cluster as determined using an empirical SFG template \citep{kir12} and the \citet{mur11a} relation.  Once simulated sources are inserted, source extraction is performed as described in Section~\ref{sec:pacs} using the full IRAC prior list plus the known positions of the simulated sources.  A simulated source is considered recovered if it is detected at $\geq2\sigma$.  We additionally split our simulated sources into radial bins from the center of the map in order to test how the completeness varies as a function of radius.  The differential completeness for the central, uniform  5$^{\prime}$x5$^{\prime}$ of each map as a function of the SFR corresponding to the input flux density of the simulated source at the redshift of the cluster can be seen in Figure~\ref{fig:completeness}.  The dashed (dotted) lines shows that 10 of the cluster maps are $\geq50\%$ ($\geq70\%$) complete at a SFR$\sim\!80 \,\Msun$ yr$^{-1}$ (SFR$\sim\!100 \,\Msun$ yr$^{-1}$).   The completeness of all maps drops by $10-15\%$ outside the uniform coverage, out to a radius of 4 arcminutes from the center of the map.  Separate completeness functions are shown for ID6 and ID10 as they share a map but the clusters are at different redshifts.  At the same SFR, ID6 is $\sim10-20\%$ more complete than ID10 due to the depth of the ID6/ID10 map.

Due to our source extraction being based on priors, we consider sources detected at a lower S/N than we would using blind source extraction.  Following \citet{mag13},  we use our Monte Carlo simulation to test the photometric accuracy and uncertainty estimates of simulated sources inserted into the map.  Photometric accuracy is defined as the standard deviation of ($S_{out}-S_{in}) / S_{out}$, where $S_{in}$ is the known input flux of the simulated sourced and $S_{out}$ is the flux recovered.  As our simulated sources are inserted into the real signal map, this test accounts for all sources of noise including confusion.  We find that our photometric accuracy is generally better than $31\%$, consistent with most of our simulated sources being recovered at $\geq3\sigma$, and with better than $50\%$ photometric accuracy for sources recovered at $\sim2\sigma$.  In addition, we examine the quantity $S_{out}-S_{in} / \sigma_s$, where $\sigma_s$ is the uncertainty on the flux density measured from the residual maps.  We find that the distribution of this quantity is a Gaussian with mean zero and a dispersion of one, indicating that our source extraction does not underestimate the uncertainties associated with a given source.

\begin{deluxetable}{cccc}
\tabletypesize{\footnotesize}
\tablecolumns{4}
\tablewidth{0pt}
\tablecaption{Summary of {\it Herschel}/PACS Imaging\label{tbl:a}}
\tablehead{
	\colhead{Cluster ID} & 
     \colhead{Short ID} &
	\colhead{RMS Sensitivity [mJy]\tablenotemark{a}} &
     \colhead{OBSIDs} }
\startdata
ISCS J1432.4+3332 & ID1 & 2.2 & 1342257535   \\
& & & 1342257536 \\ 
ISCS J1434.5+3427 & ID2 & 1.3 & 1342257748   \\
& & & 1342257749 \\
ISCS J1429.3+3437 & ID3 &  1.5 & 1342247404  \\
& &  & 1342247405 \\
ISCS J1432.6+3436 & ID4 & 1.4 & 1342257958   \\
 & & & 1342257959 \\
ISCS J1434.7+3519 & ID5 & 1.2 & 1342257962  \\
& & & 1342257963 \\
ISCS J1432.3+3253\tablenotemark{b} & ID6 & 0.9 & 1342257957   \\
& & & 1342258437 \\
ISCS J1425.3+3250 & ID7 & 1.1 & 1342248735  \\
 & & & 1342257712 \\
ISCS J1438.1+3414 & ID8 & 1.2 & 1342257746  \\
& & & 1342257747 \\
ISCS J1431.1+3459 & ID9 & 1.0 & 1342257960 \\
& & & 1342257961 \\
ISCS J1432.4+3250\tablenotemark{b} & ID10 & 0.9 & 1342257957  \\
& & & 1342258437 \\
ISCS J1426.5+3508 & ID11 & 0.6 & 1342257709 \\
 & & & 1342257710 \\
& & & 1342257711 \\
 & & & 1342248734  \\
\enddata
\tablenotetext{a}{RMS sensitivities measured in the central 5$^{\prime}$x5$^{\prime}$ region.}
\tablenotetext{b}{ID6 and ID10 were observed with the same AORs.}
\end{deluxetable} 

Selecting sources to a lower S/N may also introduce spurious detections.  Since we are using priors, this should be minimized, however, we test the random occurrence of $2\sigma$ peaks in our map by performing source extraction on randomized priors.  We find a $\sim7\%$ occurrence of a $2\sigma$ peak given random priors, consistent with the Gaussian noise expectation of $5\%$ plus a confusion noise component.    Finally, we visually inspect all PACS-detected cluster members for blending with neighboring IR sources.  We remove 10$\%$ of cluster members from our analysis due to blending.  These Monte Carlo simulations and tests provide confidence that we are able to extract sources using IRAC priors and accurately measure their flux densities and uncertainties for sources detected at $\geq2\sigma$.  

\section{Photometric Redshift Uncertainties: Pair Statistics}
\label{appendix:b}

\begin{figure}[!ht]
\makebox[\columnwidth]{\includegraphics[angle=270, trim=0 20mm 0 0, clip, scale=0.4]{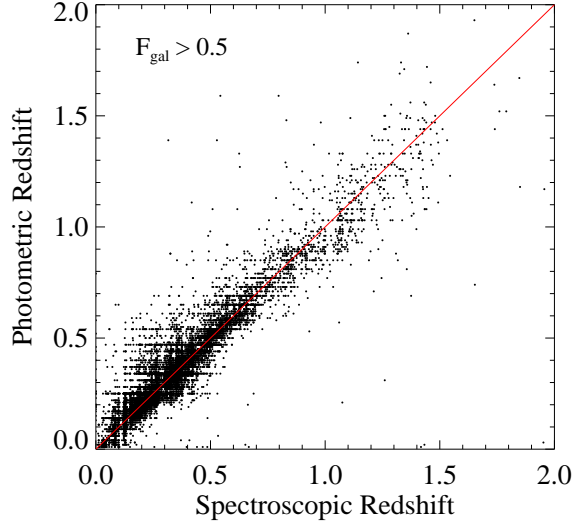}}
\caption{\footnotesize{Comparison of spectroscopic and photometric redshifts for galaxies (F$_{\rm{gal}}>0.5$) in the photometric redshift catalog.   After 5$\%$ outlier rejection, we find a photometric redshift uncertainty of $\sigma$/(1+$z$) = 0.040. The red line represents a one-to-one relation.}}
\label{fig:specz}
\end{figure}

Photometric redshift uncertainties are typically measured through comparisons with spectroscopic redshifts.   Splitting the photometric redshift catalog into unambiguous galaxy and AGN subsets, \citet{chu14} reported redshift dispersions of $\sigma$/(1+$z$) = 0.040 for galaxies and $\sigma$/(1+$z$) = 0.169 for AGN, with 5$\%$ outlier rejection.  Here we expand this comparison in order to quantify the photometric redshift uncertainties for all sources, including composites.  We match high quality spectroscopic redshifts to IRAC sources with a measured photometric redshift within 1$^{\prime\prime}$ and compare spectroscopic and photometric redshifts.  We find that the uncertainty for galaxies and galaxy composites (F$_{\rm{gal}}>0.5$) is $\sigma$/(1+$z$) = 0.040 (Figure~\ref{fig:specz}), consistent with \citet{chu14}, while for AGN and AGN composites (F$_{\rm{gal}}<0.5$), we measure $\sigma$/(1+$z$) = 0.214.

\begin{figure*}[!ht]
\centering
\includegraphics[angle=270, trim=0 10mm 0 0, clip, scale=0.5]{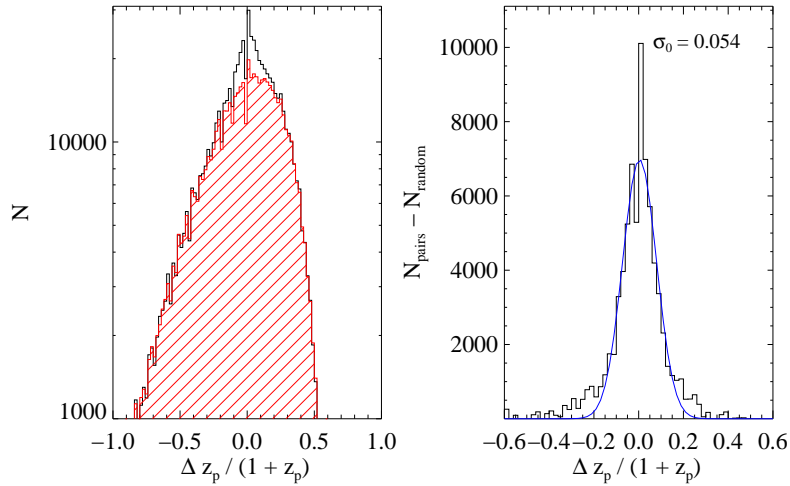}
\caption{\footnotesize{Left: The distribution of $\Delta z_p/(1+z_p)$ for close galaxy pairs ($r<30^{\prime\prime}$, black histogram) and for a random distribution (red histogram), where $z_p$ is the photometric redshift.  Right:  The residual excess from subtracting the random distribution from the distribution of galaxy pairs.  The blue line is a Gaussian fit.  The width of the Gaussian, divided by $\sqrt{2}$ to correct for double counting, gives the photometric redshift uncertainty for these sources, which is measured to be $\sigma$/(1+$z$) = 0.054. }}
\label{fig:pairs}
\end{figure*}

Though the above results indicate accurate photometric redshifts for galaxies and galaxy composites, which we expect to dominate our cluster members, we note that our spectroscopic redshift sample for non-AGN is sparse at the redshifts of interest ($1<z<1.8$).  Therefore we show here the results of an alternative method for measuring photometric redshift uncertainties: pair statistics \citep{qua10, hua13, dah13}.  Pair statistics takes advantage of the fact that some fraction of galaxy pairs with small angular separations will actually be physically associated (i.e. at the same redshift), in excess of a random distribution of projected pairs.  Figure~\ref{fig:pairs} (left) shows the distribution of $\Delta z_p/(1+z_p)$ for pairs of galaxies (F$_{\rm{gal}}>0.5$; black histogram) within 30$^{\prime\prime}$ of each other, where $\Delta z_p$ is the difference in their photometric redshifts.  This is compared to a random distribution (red histogram) where the same set of photometric redshifts are assigned random positions over the same area.  The resulting excess (right) was fit with a Gaussian distribution and the standard deviation was measured (and divided by $\sqrt{2}$ to remove double-counting).  Using this technique, we measure $\sigma$/(1+$z$) = 0.054 for all F$_{\rm{gal}}>0.5$ photometric redshifts.  To check that the photometric redshift uncertainties do not degrade as a function of redshift, we further split the photometric redshift catalog into broad redshift bins and repeat this analysis.  We find that the uncertainties are stable up to $z\sim2$.

\end{document}